\documentclass[pre,twocolumn,preprintnumbers,amsmath,amssymb,showpacs,
 nofootinbib,floatfix]{revtex4}

\usepackage{graphicx,bm}

\makeatletter
\def\graphicscale{\twocolumn@sw{0.3}{0.4}}
\def\graphicthreescale{\twocolumn@sw{0.3}{0.4}}

\makeatother

\def\spose#1{\hbox to 0pt{#1\hss}}
\def\lesssim{\mathrel{\spose{\lower 3pt\hbox{$\mathchar"218$}}
 \raise 2.0pt\hbox{$\mathchar"13C$}}}
\def\gtrsim{\mathrel{\spose{\lower 3pt\hbox{$\mathchar"218$}}
 \raise 2.0pt\hbox{$\mathchar"13E$}}}

\def\<{\langle}
\def\>{\rangle}

\begin{document}

\title{                                                                           
Renormalization-group flow and asymptotic behaviors at the 
Berezinskii-Kosterlitz-Thouless transitions
}

\author{Andrea Pelissetto$^1$ and Ettore Vicari$^2$}
\address{$^1$ Dipartimento di Fisica dell'Universit\`a di Roma ``La Sapienza"
        and INFN, Sezione di Roma I, I-00185 Roma, Italy}
\address{$^2$ Dipartimento di Fisica dell'Universit\`a di Pisa
        and INFN, Sezione di Pisa, I-56127 Pisa, Italy}

\begin{abstract}
We investigate the general features of the renormalization-group flow
at the Berezinskii-Kosterlitz-Thouless (BKT) transition, providing a
thorough quantitative description of the asymptotc critical behavior,
including the multiplicative and subleading logarithmic corrections.
For this purpose, we consider the RG flow of the sine-Gordon model
around the renormalizable point which describes the BKT transition.
We reduce the corresponding $\beta$-functions to a universal canonical
form, valid to all perturbative orders. Then, we determine the
asymptotic solutions of the RG equations in various critical regimes:
the infinite-volume critical behavior in the disordered phase, the
finite-size scaling limit for homogeneous systems of finite size, and
the trap-size scaling limit occurring in 2D bosonic particle systems
trapped by an external space-dependent potential.
\end{abstract}

\pacs{05.10.Cc, 05.70.Jk, 67.25.dj, 67.85.-d, 74.78.-w}                           
%05.10.Cc       Renormalization group methods 
%05.70.Jk       Critical point phenomena    
%67.25.dj       Superfluid transition and critical phenomena
%67.85.-d       Ultracold gases trapped gases
%74.78.-w       Superconducting films and low-dimensional structures  

%74.81.Fa Josephson junction arrays                                               
%75.10.Hk classical spin models                                                   
%05.10.Ln       Monte Carlo methods                                               
%05.30.Jp       Boson systems, 
%64.60.F-       Equilibrium properties near critical points, critical exponents 
%64.60.an 	Finite-size systems
%64.60.fd       General theory of critical region behavior

\maketitle

\section{Introduction}

The Berezinskii-Kosterlitz-Thouless (BKT)
theory~\cite{KT-73,B-72,Kosterlitz-74,JKKN-77} describes
finite-temperature transitions in two-dimensional (2D) systems with a
global U(1) symmetry, which belong to the so-called 2D XY universality
class.  BKT transitions are quite peculiar, since the low-temperature
phase is not characterized by long-range order and the emergence of a
nonvanishing order parameter~\cite{MW-66,H-67}, but rather by
quasi-long range order (QLRO) with correlations decaying algebraically
at large distance. For example, a 2D fluid of identical bosons cannot
undergo Bose-Einstein condensation.  Above $T_c$ these systems show a
standard disordered phase with exponentially decaying correlations.
The BKT theory predicts an exponential increase of the correlation
length when approaching the transition point $T_c$ from above, as
$\xi\sim \exp(c /\sqrt{\tau})$ with $\tau\equiv T/T_c-1$.  BKT
transitions are generally expected in 2D systems of interacting
bosonic atoms, such as those investigated in experiments with trapped
atomic gases~\cite{HKCBD-06,KHD-07,HKCRD-08,CRRHP-09,HZGC-10}, in
liquid helium films~\cite{GKMD-08}, in arrays of Josephson
junctions~\cite{RGBSN-81}, etc.  These experiments have provided
evidence of the general features predicted by the BKT theory.

A standard representative model of the 2D XY universality class is the
classical 2D XY model. Its Hamiltonian is
\begin{equation}
H_{\rm XY} = - J \sum_{\langle ij \rangle } {\rm Re}\, 
\bar{\psi}_i \psi_j, \qquad \psi_i \in {\rm U}(1),
\label{XYmodel}
\end{equation}
where the sum runs over the bonds of a square lattice.  Its phase
diagram shows a BKT transition between a high-temperature disordered
phase and a low-temperature QLRO phase, at~\cite{HP-97,Hasenbusch-05,KO-12}
$J \beta_c=1.1199(1)$.  The asymptotic behaviors at the BKT
transition, and in particular their logarithmic corrections, have been
much investigated, see e.g.
Refs.~\cite{BC-93-94,KI-95,KI-96,CPRV-96,Janke-97,KI-97,JH-98,Balog-01,BNNPSW-01,%
CS-03,Hasenbusch-05,BP-08,Arisue-09,KO-12}, by analytical and
numerical studies.

In this paper we investigate the general features of the
renormalization-group (RG) flow at the BKT transition, providing a
complete characterization of the asymptotic BKT behaviors,
and in particular of the multiplicative and subleading logarithmic
corrections.  For this purpose, we exploit the mapping between the 2D XY or
Coulomb-gas models and the sine-Gordon (SG) model, whose RG flow
around the renormalizable point describes the BKT
transition~\cite{AGG-80}.  In order to investigate its RG flow, we
first reduce the SG $\beta$-functions to a canonical universal
form. Explicitly, we show that we can define appropriate couplings $u$
and $v$ so that the associated $\beta$-functions have the form
$\beta_u=-uv$ and $\beta_v=-u^2[1+v f(v^2)]$ to all orders of the
perturbative expansion in powers of $u$ and $v$.  The universal
function $f(v^2)$ cannot be determined by general arguments, but only
by means of detailed calculations in the SG model (at present only
$f(0)$ is known).  Then, generalizing the RG study of
Ref.~\cite{Balog-01}, we determine the asymptotic solutions of the RG
equations in some different critical regimes: the infinite-volume
critical behavior in the disordered phase, the finite-size scaling
(FSS) limit for homogeneous systems of finite
size~\cite{FBJ-73,Cardy-88}, and the trap-size scaling (TSS)
limit~\cite{CV-09} in 2D bosonic particle systems trapped by an
external space-dependent potential.  The latter results are relevant
for experimental investigations of trapped quasi-2D atomic gases, such
as those reported in
Refs.~\cite{HKCBD-06,KHD-07,HKCRD-08,CRRHP-09,HZGC-10}.

The paper is organized as follows.  In Sec.~\ref{RGideas} we summarize
some of the RG ideas that we use to analyze the BKT RG flow.
Sec.~\ref{canform} reports the derivation of the canonical form of the
$\beta$-functions and outlines the main features of the RG flow which
they describe.  In Sec.~\ref{infvolsec} we derive the asymptotic
critical behavior of some observables, such as the correlation
function $\langle\bar{\psi}(x)\psi(y)\rangle$ of the XY model and its
low-momentum components, when approaching the critical point from the
high-temperature phase.  In Sec.~\ref{fsssec} we discuss the
asymptotic behavior in the FSS limit, i.e. in the infinite-volume
limit keeping the ratio between the system size and the correlation
length fixed, and in particular at $T_c$. Multiplicative logarithms
appear in the two-point function and in the related susceptibility.
Sec.~\ref{TSSsec} is devoted to an analysis of the TSS behavior at the
BKT transition. We show that, at $T_c$, the critical behavior with
respect to the trap size must include new multiplicative logarithms.
Finally, in Sec.~\ref{secco} we draw our conclusions and summarize the
main results of the paper.  The various appendices report some
technical details of the derivations of the results.

\section{A short summary of the renormalization-group ideas} 
\label{RGideas}

Before discussing the RG flow at the BKT transition, we would like to
recall a number of ideas concerning the RG flow, which we then apply
to the study of the BKT critical behavior.  We consider a generic
critical system depending on a set of scaling fields $u_{10}$,
$u_{20}$, and so on.  The RG flow with respect to a length rescaling
$b$ is defined by \cite{Wegner-76}
\begin{equation}
    b {{\rm d} u_i\over {\rm d}b} = \beta_i(u_1,u_2,\ldots),
\end{equation}
with boundary condition $u_i(b=1) = u_{i0}$. Then, the scaling part of
the free-energy density satisfies
\begin{equation}
{\cal F} ( u_{10},u_{20},\ldots) = 
   b^{-d}{\cal F}[u_1(b),u_2(b),\ldots],
\label{scaling-F}
\end{equation}
where $d$ is the space dimension.  Analogously, the correlation length
satisfies
\begin{equation}
{\xi} ( u_{10},u_{20},\ldots) =
   b {\xi} [u_1(b),u_2(b),\ldots].
\label{RG-xigen}
\end{equation}
An operator that renormalizes multiplicatively scales as
\begin{eqnarray}
&&{\cal O} ( u_{10},u_{20},\ldots) = \\
&&=
   b^{d_{\cal O}} Z_{\cal O}[u_1(b),u_2(b),\ldots]
   {\cal O} [u_1(b),u_2(b),\ldots], 
\nonumber
\end{eqnarray}
where $d_{\cal O}$ is its power-counting dimension and $Z_{\cal O}$
satisfies the RG equation
\begin{equation}
    b {{\rm d}\ln Z_{\cal O}\over {\rm d}b} = \gamma_{\cal O}(u_1,u_2,\ldots),
\qquad Z_{\cal O}(b=1) = 1,
\label{RG-Zeq}
\end{equation}
where $\gamma_{\cal O}$ is
the anomalous dimension associated with $\cal O$.

The RG flow can be characterized in terms of characteristic surfaces.
For a given set of initial conditions we consider the functions
$f_i(u_1)$ for $i\ge 2$ (the choice of the first scaling field to
parametrize the flow is arbitrary; any other choice would work equally
well), which are solutions of the equations
\begin{equation}
{{\rm d} f_i\over {\rm d} u_1} 
   = {\beta_i[u_1,f_2(u_1),f_3(u_1),\ldots] \over 
      \beta_1[u_1,f_2(u_1),f_3(u_1),\ldots] } 
\end{equation}
with initial conditions $f_i(u_{10}) = u_{i0}$. The hypersurface $H_i$
of equation
$u_i = f_i(u_1)$ is invariant under the RG flow, since 
\begin{equation}
b {{\rm d} \over {\rm d}b} [u_i(b) - f_i(u_1(b))] = 
   \beta_i - {\beta_i\over \beta_1} \beta_1 = 0.
\end{equation}
Therefore the flow line lies in the intersection of all the $H_i$
hypersurfaces.

We can also take into account the size $L$ of the system.  With the
usual hypotheses \cite{PHA-91,Privman-90,SS-99}
of the FSS theory, this is
achieved by adding a term $L/b$ in the scaling Ansatz:
\begin{equation}
{\cal F}( u_{10},u_{20},\ldots,L) = 
   b^{-d}{\cal F} [u_1(b),u_2(b),\ldots,L/b].
\label{scaling-F-L}
\end{equation}
Also correlation functions that depend on coordinates $x$, $y$, etc. 
can be analyzed. In this case the RG mapping is simply $x\to x/b$, 
$y \to y/b$, etc. 

In the following we shall use the notation
\begin{equation}
l \equiv \ln b,\qquad 
{{\rm d} \over {\rm d}l} = b {{\rm  d} \over {\rm d}b}  .
\label{defl}
\end{equation}

\section{Canonical form of the BKT beta functions
and renormalization-group flow}
\label{canform}

We want to study the general features of the RG flow at the BKT
transition of 2D systems with U(1) symmetry.  For this purpose, we
consider the SG field-theoretical model, see, e.g.,
Ref.~\cite{ZJ-book}, defined by the Lagrangian
 \begin{equation}
 {\cal L}_{\rm SG} = {1\over 2}(\partial_\mu\phi)^2
 + {\alpha\over a^2 \beta^2}\left[ 1 - {\rm cos}(\beta\phi)\right],
  \label{sineG}
 \end{equation}
 where $\alpha$ and $\beta$ are dimensionless coupling constants, and
 $a$ is an ultraviolet length scale.  The RG flow around the fixed
 point $\beta^*=\sqrt{8\pi}$, $\alpha^*=0$ describes the BKT
 transition~\cite{AGG-80}. It can be investigated by a renormalizable
 two-parameter perturbative expansion in powers of $\delta$, defined
 by $\beta^2=8\pi(1+\delta)$, and $\alpha$.  Their $\beta$ functions
 have been computed to two-loop order~\cite{AGG-80,BH-00},
 obtaining~\cite{footnotesbf}
\begin{eqnarray}
&&\beta_\alpha = -2\alpha\delta - {5\over 64} \alpha^3,\label{betaf1}\\
&&\beta_\delta = -{1\over 32} \alpha^2 + {1\over 16} \alpha^2\delta.
\label{betaf2}
\end{eqnarray}

Under an appropriate analytic redefinition of the couplings $\alpha$
and $\delta$, the above two-loop $\beta$-functions can be rewritten as 
\begin{eqnarray}
&&\beta_u = - uv, \label{twolred}\\
&&\beta_v = - u^2 - {3\over 2} u^2 v.
\nonumber
\end{eqnarray}
The coefficient $-3/2$ is universal in the following sense: there is no
redefinition of the couplings $u' = U(u,v)$ and $v' = V(u,v)$ such that 
$\beta_{u'} = - u' v'$ and $\beta_{v'} = - {u'}^2  - c {u'}^2 v' + \ldots$,
with $c \not = -3/2$. 

This reduction to a universal form can be extended to all orders,
leading to the most general canonical form of the $\beta$-functions
which is compatible with the invariance of the model under $\alpha\to
-\alpha$ (it corresponds to a shift of $\pi/\beta$ in the field $\phi$).
We prove that, by an analytic redefinition of the couplings
\begin{eqnarray}
\alpha &=& a_{\alpha,10} u +
   \sum_{n+m\ge 2} a_{\alpha,nm} u^n v^m, \\
\delta &=& a_{\delta,10} v +
   \sum_{n+m\ge 2} a_{\delta,nm} u^n v^m,
\end{eqnarray}
the $\beta$-functions of the SG model can be rewritten in the 
general form
\begin{eqnarray}
\beta_u(u,v) &=& - u v, 
\label{betau}\\
\beta_v(u,v) &=& - u^2 [1 + v f(v^2)],
\label{betav-1} 
\end{eqnarray}
where $f(x)$ has an expansion of the form
\begin{equation}
f(x) = b_0 + b_1 x + b_2 x^2 + \ldots
\label{fvdef}
\end{equation}
This representation of the $\beta$-functions is universal, in the sense 
that, by redefining the couplings, it is not possible to obtain
$\beta$ functions of the same form, i.e., $\beta_{u'} = - u'v'$ and 
$\beta_{v'} = - {u'}^2  [1 + v' g({v'}^2)]$,  with $g(x) \not= f(x)$.
The proof is outlined in
App.~\ref{app:beta} (this result was already conjectured in
Ref.~\cite{APV-10} without proof).  The coefficient $b_0$ can be read off
from Eq.~(\ref{twolred}),
\begin{equation}
b_0 = {3/2}, 
\label{b1res}
\end{equation}
while the higher-order terms, $b_i$, $i\ge 1$, are unknown.
Note that the evaluation of the next universal
coefficient $b_1$ requires a perturbative calculation of the SG
$\beta$-functions to four loops.

The analysis of the RG flow can be performed following the method
outlined in Sec.~\ref{RGideas}, see also
Refs.~\cite{AGG-80,Balog-01}. First, we define the RG invariant
function
\begin{eqnarray}
Q(u,v)= u^2 - F(v),
\label{qdef} 
\end{eqnarray}
where
\begin{eqnarray}
F(v) &=& 2 \int_0^v {w dw\over 1 + w f(w^2)} \label{ffv}\\
&=& v^2 - {2 b_0\over 3} v^3 + 
{b_0^2\over 2} v^4 + O(v^5),
\nonumber
\end{eqnarray}
which satisfies 
\begin{equation}
{dQ\over dl} = {\partial Q\over \partial u} \beta_u(u,v) + 
   {\partial Q\over \partial v} \beta_v(u,v)  = 0,
\end{equation}
where $l$ is the flow parameter. The RG flow follows the lines $Q =\,
$constant.  It is thus natural to parametrize the RG flow in terms of
$Q$ and $v(l)$.  Since
\begin{equation}
{dv\over dl} = \beta_v(u,v) = - [Q + F(v)][1 + v f(v^2)],
\label{betavqf}
\end{equation}
we obtain 
\begin{equation}
l = - \int_{v_0}^{v(l)} {dw\over [Q + F(w)][1 + w f(w^2)]},
\label{flow-vl-basic}
\end{equation}
where $v(l=0) = v_0$.

It is important to stress that Eqs.~(\ref{betau}) and (\ref{betav-1})
are the two $\beta$-functions associated with the marginal operators
that characterize the BKT transition.  For a full understanding of the
scaling corrections one should also consider the contributions of the
subleading operators, which are suppressed by powers of the critical
length scale.  The most relevant subleading operator at the BKT
transition is expected to have RG dimension $-2$, as in the Gaussian
spin-wave theory, see, e.g., Ref.~\cite{HP-97}.  In the standard RG
language this corresponds to a scaling-correction exponent $\omega=2$.
Thus, subleading operators induce corrections of order $\xi^{-2}$ in
the high-temperature infinite-volume limit and of order $L^{-2}$ in
the FSS limit (apart from corrections arising from the boundary
conditions~\cite{Diehl-86,Hasenbusch-12}, which are expected to be $O(L^{-1})$; they
are absent in the case of boundary conditions preserving translation
invariance, such as periodic boundary conditions).  Additional
multiplicative logarithms may also appear (hence, corrections might
scale as $\xi^{-2}\ln^p\xi$, $L^{-2}\ln^q L$), because of the possible
resonance between the subleading and the marginal operators
\cite{Wegner-76}, the difference between their RG dimensions being an
integer number.  In the following we shall not consider these scaling
corrections, since our focus will be mainly on the logarithmic
corrections to the leading behavior that can be obtained by
considering only the two marginal couplings.

\section{Infinite-volume results at the BKT transition}
\label{infvolsec}

\begin{figure}[tbp]
\includegraphics*[scale=\graphicscale]{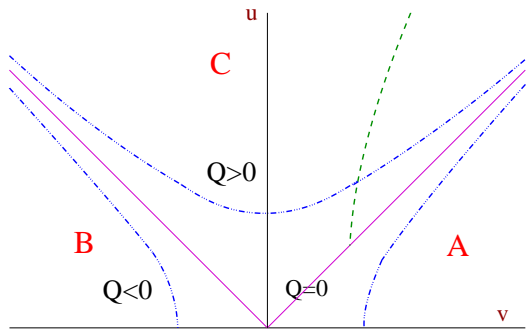}
\caption{Sketch of the RG flow at the BKT transition.  The dashed
curve in the region $C$ ($v>0$) shows the approach to criticality
($Q=0$ line) of the 2D XY model from the HT phase.  The correlation
length is singular along the $Q=0$ line in the region $v>0$, while it
is analytical along the $Q=0$ for $v<0$.  }
\label{rgflow}
\end{figure}

Let us now apply these results to the XY model.  In Fig.~\ref{rgflow}
we show a sketch of the RG flow. Repeating the discussion of
Refs.~\cite{Kosterlitz-74,JKKN-77,Balog-01} the XY model can be mapped
onto a line in the $(u,v)$ plane with $v > 0$.  The BKT transition is
the intersection of this line with the line $Q=0$ and the
high-temperature phase corresponds to $Q > 0$ (region C with $v>0$ in
Fig.~\ref{rgflow}).  Thus, $Q$ plays the role of thermal nonlinear
scaling field, i.e.
\begin{equation}
Q = q_1 \tau + q_2 \tau^2 + \ldots
\end{equation}
where $\tau = T/T_c-1$, and $q_i$ are nonuniversal coefficients.  In
the following we shall use $Q$ instead of $\tau$.

Let us first consider the infinite-volume correlation length
$\xi_\infty(\tau) \equiv \xi_\infty(Q,v)$, which we may define by
using the second-moment of the two-point correlation function
\begin{equation}
G(x,y)\equiv \langle \bar{\psi}(x) \psi(y)\rangle,
\label{g2def}
\end{equation}
or its large-distance exponential decay.  For all $v > 0$ the
correlation length is singular as $Q\to 0$. On the other hand, for $v
< 0$, the correlation length is analytic on the line $Q = 0$ (see
Ref.~\cite{Balog-01}).  Hence, in order to obtain the singular
behavior, we use the RG equations to flow from the starting point $v_0
> 0$ to the negative point $v = -1$.

We determine $l_0$ by requiring
\begin{equation}
   v(l_0) = -1,
\end{equation}
so that 
\begin{equation}
l_0 = \int^{v_0}_{-1} {dw\over [Q + F(w)][1 + w f(w^2)]} = I(Q,v_0).
\end{equation}
Then, Eq.~(\ref{RG-xigen}) gives
\begin{equation}
\xi_\infty(Q,v_0) = e^{l} \xi_\infty(Q,v(l)) = e^{I(Q,v_0)} \xi_\infty(Q,-1).
\label{infvol-xiQv0}
\end{equation}
Since $\xi_\infty(Q,-1)$ is analytic in $Q$, the singular part is given by
the exponential term.  The behavior of $\xi_\infty(\tau)$ for $\tau \to 0$ is
obtained by computing the asymptotic expansion of $I(Q,v_0)$ for $Q\to
0$, which can be written in the form
\begin{equation}
I(Q,v_0) = {1\over \sqrt{Q}} \sum_{n=0} I_n Q^n + 
            \sum_{n=0} Y_{n}(v_0) Q^n ,
\end{equation}
see App.~\ref{App.B} for its derivation.
The nonanalytic terms in this expansion depend only of the coefficients
$b_k$ which appear in Eq.~(\ref{betav-1}).  The first two
coefficients are
\begin{eqnarray}
I_0 &=& \pi, \nonumber \\
I_1 &=& -{\pi b_0^2\over 12} = -{3 \pi\over 16} .
\end{eqnarray}
Correspondingly, we obtain
\begin{equation}
\xi_\infty(\tau) = X \exp (\pi/\sqrt{Q}) [1 + I_1 \sqrt{Q} + O(Q)].
\label{xiKT-Q}
\end{equation}
Expanding $Q$ in powers of $\tau$ we obtain the celebrated BKT
expression for the correlation length~\cite{KT-73,Kosterlitz-74}.  Of
course, this is also consistent with the exact result for the
correlation length of the F-model, computed using transfer-matrix
techniques, see e.g. \cite{Baxter-82,HP-97}.

These results allow us to derive the critical behavior of of generic
RG invariant dimensionless quantities, which we generically denote by
$R$.  Examples of such quantities are ratios of length scales
and the so-called zero-momentum $2n$-point renormalized couplings
(related to the small-magnetixation expansion of the scaling free energy),
see e.g. Refs.~\cite{CPRV-96,CPRV-96b,PV-98,BNNPSW-00,PV-r}.
Indeed, $R$ satisfies the scaling relation
\begin{equation}
R(Q,v_0) = R[Q,v(l)],
\end{equation}
for any $l$. This implies that $R(Q,v_0)$ is independent of $v(l)$,
hence an analytic function of $Q$ and therefore of $\tau$. It follows
\begin{equation}
R(\tau) = R^* + {c_R\over\ln^2(\xi_\infty/X)} + O(\ln^{-3}\xi_\infty),
\label{R-infvol}
\end{equation}
where the costant $c_R$ is expected to be universal.

Let us now consider a generic operator that renormalizes
multiplicatively.  Writing the corresponding anomalous dimension
$\gamma_{\cal O}(u,v)$ in terms of $Q$ and $v$, 
from Eq.~(\ref{RG-Zeq})  we obtain
\begin{equation}
Z_{\cal O}(Q,v_0) = \left[\exp \int_{v_0}^{-1} dw\, 
    {\gamma_{\cal O} (Q,w) \over \beta_v(Q,w)} \right] Z_{\cal
O}(Q,-1).
\label{ZO-exp}
\end{equation}
Taking into account the symmetry properties of the SG model, we write
the expansion of $\gamma_O$ as
\begin{eqnarray}
&&\gamma_{\cal O}= g_{00} + g_{01} v + g_{02} v^2 + g_{20} u^2 + 
      O(v^3,u^2 v^2, u^4)=\nonumber \\
&&= g_{00} + g_{01} v + (g_{02} + g_{20}) v^2 + g_{20} Q +
      O(v^3,Q v^2, Q^2).\nonumber\\
\end{eqnarray}
It is important to discuss the universality of this expansion.  There
is a residual transformation of the couplings that leaves the
$\beta$-functions (\ref{betau}) and (\ref{betav-1}) invariant:
\begin{eqnarray}
&&u' = u + A u v + \ldots,\\
&&v' = v + A u^2 + \ldots ,
\end{eqnarray}
for any $A$. The RG invariant function $Q$ is invariant under the
transformation and so are the coefficients $g_{00}$, $g_{01}$, and
$g_{02}$, hence they are universal. Instead, $g_{20}$ can be changed
at will, hence it is model dependent.

We can now rewrite Eq.~(\ref{ZO-exp}) as
\begin{eqnarray}
&&Z_{\cal O}(Q,v_0) = \label{z0qveq}\\
&&
e^{l_0 g_{00}} \left[\exp \int_{v_0}^{-1} dw\, 
    {\gamma_{\cal O} (Q,w) - g_{00} \over \beta_v(Q,w)} 
   \right] Z_{\cal O}(Q,-1).
\nonumber
\end{eqnarray}
Collecting everything together we obtain
\begin{equation}
{{\cal O}(Q,v) \over \xi_\infty(Q,v)^{d_{\cal O} + g_{00}} }
= C(Q)
   \exp \int_{v_0}^{-1} dw\,
    {\gamma_{\cal O} (Q,w) - g_{00} \over \beta_v(Q,w)},
\end{equation}
where $C(Q)$ is an analytic function of $Q$. For $Q\to 0$ we obtain an
expansion of the form
\begin{equation}
{{\cal O}(Q,v)\over \xi_\infty(Q,v)^{d_{\cal O} + g_{00}}} =
a_0 C(Q) [1 + a_1 \sqrt{Q} + a_2 Q + \ldots],
\end{equation}
where $a_0$ and $a_2$ depend on nonuniversal details (the starting
point $v_0$, for instance), while $a_1$ is universal, since it only
depends on the universal coefficient $g_{02}$:
\begin{equation}
a_1 = - \pi g_{02}.
\end{equation}

The above result can be specialized to the susceptibility, defined as
the space integral of the two-point function (\ref{g2def}).
Perturbation theory for the scaling dimension of the spin correlation
function gives~\cite{AGG-80}
\begin{eqnarray}
&&\gamma = -{1\over 4} + {1\over 4}\delta - {1\over 4} \delta^2 +
h_1\alpha^2+ \ldots , 
\end{eqnarray}
where $h_1$ is an unknown coefficient. If we perform the redefinitions
$(\alpha,\delta) \to (u,v)$ considered before, we can rewrite $\gamma$ as
\begin{eqnarray}
&&\gamma = -{1\over 4} + {1\over 8}v - {1\over 16} v^2 + g_{20} u^2\ldots
\label{gammapert}
\end{eqnarray}
with arbitrary $g_{20}$.  The previous results show that, in the
infinite-volume limit, the susceptibility satisfies the scaling law
\begin{equation}
\chi\xi^{-7/4}_\infty  = A(1 + c_1 \sqrt{Q} + c_2 Q + \ldots).
\end{equation}
The coefficient $c_1$ can be computed exactly. Since $g_{02} = -1/16$,
we obtain
\begin{equation}
   c_1 = {\pi\over 16}.
\end{equation}
Using Eq.~(\ref{xiKT-Q}) we can write
\begin{equation}
\sqrt{Q} = {\pi\over\ln(\xi_\infty/X)} + O(\ln^{-2}\xi_\infty)
\end{equation}
and obtain 
\begin{equation}
\chi\xi^{-7/4}_\infty  = A_\chi
   \left[1 + {\pi^2\over 16 \ln(\xi_\infty/X)} + 
         O(1/\ln^{2}\xi_\infty) \right].
\label{chi-expRG}
\end{equation}
As already noted in Ref.~\cite{Balog-01}, RG predicts the absence of 
a leading logarithmic singular term.
Note also that, at variance with what happens with RG invariant quantities,
cf. Eq.~(\ref{R-infvol}), corrections decay as $1/\ln \xi_\infty$ and not 
as $1/\ln^2 \xi_\infty$. Moreover, the coefficient of 
the leading logarithmic scaling correction is universal.

We should note that in the literature, see, e.g.,
\cite{BC-93-94,KI-95,KI-96,CPRV-96,Janke-97,JH-98}, it was often assumed 
that the correct scaling behavior is $\chi\xi^{-7/4}_\infty \sim 
(\ln \xi_\infty)^{1/8}$. Such a behavior was 
derived as follows. Since the two-point function at the critical point
scales as \cite{KT-73,Kosterlitz-74,AGG-80} $G(r)\sim (\ln r)^{1/8}/r^{1/4}$,
it was argued that 
\begin{equation}
\chi \sim \int_{r<\xi_\infty} d^2r\, G(r) \sim \xi_\infty^{7/4} (\ln
\xi_\infty)^{1/8}.
\end{equation}
However, in the presence of logarithmic corrections it is not clear if
one should cutoff the integral at $\xi_\infty$, or rather at
$\xi_\infty (\ln \xi_\infty)^q$. If we use the latter cutoff we can
freely change the leading logarithmic behavior by changing $q$. In
particular, if $q = - 1/14$, we obtain Eq.~(\ref{chi-expRG}).  
The most recent MC simulations of the XY model in the thermodynamic
limit confirm the absence of the leading singular term
\cite{BNNPSW-01}. The analyses of  
high-temperature expansions have led to apparently contradictory
results: the results Ref.~\cite{BP-08} are consistent with
Eq.~(\ref{chi-expRG}), while the analysis of Ref.~\cite{Arisue-09} 
apparently favors the presence of a leading logarithmic correction.
On the other hand, we should note that the resummation of 
high-temperature expansions
at finite order can hardly reconstruct an 
asymptotic behavior 
at the critical point with logarithmic corrections.

\section{Finite-size scaling}
\label{fsssec}

\subsection{Finite-size scaling in the high-temperature phase}
\label{fssht}

In order to study the FSS regime, we add $L/b = L e^{-l}$ in
the scaling Ansatz. If $Q\not=0$, i.e. we are not at the critical
point, we can study the FSS regime as we did in the previous
section. If we choose $l=l_0$ such that $v(l_0) = -1$, we can write
\begin{equation}
L e^{-l_0} = L e^{-I(Q,v_0)} = \xi_\infty(Q,-1) \times 
   {L\over \xi_\infty (Q,v_0)},
\end{equation}
where $\xi_\infty (Q,v_0)$ is the infinite-volume correlation length,
and $\xi_\infty(Q,-1)$ is an analytic function of $Q$, which is finite
for $Q\to 0$.  The finite-size correlation length must satisfy the
equation
\begin{equation}
\xi(Q,v_0,L) = e^{l_0} \xi(Q,-1,L e^{-l_0}).
\end{equation}
For $Q \to 0$, introducing the FSS variable
\begin{equation}
z \equiv  \xi_\infty/L, 
\label{zdef}
\end{equation}
we obtain 
\begin{eqnarray}
{\xi(Q,v_0,L)\over \xi_\infty (Q,v_0)} &=& 
   {\xi(Q,-1,L e^{-l_0})\over \xi_\infty (Q,-1)} \\
&=&
   A(z) + Q B(z) + O(Q^2).\nonumber
\end{eqnarray}
Hence, if we use $z$ as basic FSS variable, scaling corrections decay
as $1/\ln^2 \xi_\infty$. It is clear that the same arguments apply to
any observable, as long as we divide it by its infinite-volume limit.
Hence
\begin{eqnarray}
{{\cal O}(Q,v_0,L) \over {\cal O}_\infty(Q,v_0)} &=& 
{{\cal O}(Q,-1,e^{-l_0} L) \over {\cal O}_\infty(Q,-1)} \\
&=& 
A_{\cal O}(z) + Q B_{\cal O} (z) + O(Q^2).\nonumber
\end{eqnarray}
All these relations hold as long as $z$ is finite.  The
infinite-volume limit is not uniform in $z$ and indeed the scaling
functions $A_{\cal O}(z)$, $B_{\cal O} (z)$, etc, are singular for
$z\to \infty$.  The limiting behavior for $z\to \infty$, i.e. the
asymptotic behavior at the critical point will be discussed in the
next section.

\subsection{Finite-size behavior at the critical point}
\label{cpfss}

The finite-size behavior at $T_c$ is not simply obtained by extending
the results of Sec.~\ref{fssht} 
to $T_c$.  We consider the correlation length in a
finite box of size $L$, $\xi(Q,v,L)$. For finite $L$, we can take 
the limit $Q\to 0$ and obtain $\xi(0,v,L)$, which is singular as $L\to
\infty$. At $T=T_c$ it is convenient to fix $l = \ln L$, which gives
\begin{equation}
\xi(0,v,L) = e^l \xi[0,v(l),e^{-l}L] = L \xi[0,v(L),1],
\end{equation}
where $v(L) = v(l=\ln L)$.
The function $\xi[0,v(L),1]$ is analytic for any $L$, since the size is 
equal to 1, hence we have 
\begin{equation}
\xi(0,v,L) = L [s_0 + s_1 v(L) + s_2 v(L)^2 + \ldots],
\end{equation}
where the coefficients $s_i$ are universal and $v(L)$ is defined by
\begin{equation}
\ln L = \int^{v_0}_{v(L)} {dw\over F(w)[1 + w f(w^2)]}.
\end{equation}
For $L\to \infty$, the effective coupling $v(L)$ vanishes as $1/\ln L$. 
In this limit we obtain
\begin{eqnarray}
&&\ln L = {1\over v(L)} + {b_0\over 3} \ln v(L) + K \label{lnL_flow}\\ 
&& - 
   \int_{0}^{v(L)} dw \left\{{1\over F(w)[1 + w f(w^2)]} - {1\over w^2} + 
     {b_0\over 3 w} \right\},
\nonumber
\end{eqnarray}
where 
\begin{equation}
   K = \int_{0}^{v_0} dw \left\{{1\over F(w)[1 + w f(w^2)]} - {1\over w^2} + 
     {b_0\over 3 w} \right\}.
\label{defK}
\end{equation}
In Eq.~(\ref{lnL_flow}) all terms are universal, except for the
constant $K$, which encodes all microscopic details. If we define
\begin{equation}
\mu \equiv  \ln (Le^{-K}), 
\label{mudef}
\end{equation}
for $v(L)\to 0$ we obtain the expansion
\begin{equation}
\mu = {1\over v(L)} + {b_0\over 3} \ln v(L) + {5 b_0^2\over 18} v(L) + 
    O(v^2).
\label{muexp-FSS}
\end{equation}
The general solution is
\begin{eqnarray}
v(L) &=& {1\over \mu} + \sigma_1 {\ln \mu\over \mu^2} + 
   \sigma_2 {\ln^2 \mu\over \mu^3} + \label{vlsol}\\ 
&+&  \sigma_3 {\ln \mu\over \mu^3} + 
   \sigma_4 {1\over \mu^3} + O(\mu^{-4} \ln^3\mu),
\nonumber 
\end{eqnarray}
where all coefficients of the expansion are universal, and $\sigma_i$
up to $i=4$ can be computed in terms of $b_0$ only.  The terms of
order $\ln^{n-1}\mu/\mu^n$ can be resummed, obtaining
\begin{equation}
v(L) = {1\over \mu + {1\over 2} \ln \mu} + O(\mu^{-3}\ln\mu).
\label{vlexp}
\end{equation}
Indeed, we can rewrite Eq.~(\ref{muexp-FSS}) as
\begin{equation}
{1\over v} \approx \mu - {b_0\over 3} \ln v \approx 
          \mu + {b_0\over 3} \ln \mu = \mu + {1\over2} \ln \mu.
\end{equation}
The above results allow us to derive the asymptotic finite-size behavior
at $T_c$ of generic RG invariant dimensionless quantities $R$,  such as ratios
$\xi/L$ for any definition of length scale, Binder cumulants and
the helicity modulus $\Upsilon$. 
They are expected to behave as
\begin{equation}
R(L) =  R^* + C_R  v(L) + O(v^2)
\label{rexp}
\end{equation}
where the $R^*$ and $C_R$ are universal, although they may depend
on the shape of the finite volume and the boundary conditions.
Then, using Eq.~(\ref{vlexp}), we obtain
\begin{equation}
R \approx R^* + {C_R\over   \mu + {1\over 2} \ln \mu} + O(\mu^{-2}).
\label{rrstar}
\end{equation}
This result improves earlier asymptotic expansions, see, e.g.,
Refs.~\cite{Hasenbusch-05,HPV-05,Hasenbusch-08,Hasenbusch-09}.  The
asymptotic values $R^*$ and $C_R$ can be computed within the spin-wave
theory, as shown in Refs.~\cite{Hasenbusch-05,Hasenbusch-08}.  For
example, in the case of the helicity modulus in a square lattice with
periodic boundary conditions, $\Upsilon^*=0.636508...$ and
$C_\Upsilon=0.318899...$~\cite{Hasenbusch-05}.  
An analogous result applies to the exponential correlation length
$\xi_e$ in a strip $L\times\infty$, i.e. for $R_{\xi_e}\equiv \xi_e/L$,
with $R_{\xi_e}^*=4/\pi$ and $C_{\xi_e}=2/\pi$~\cite{Hasenbusch-05}.

It is worth noting that the behavior of RG invariant quantities at
$T_c$ has different features with respect to the infinite-volume case.
Here corrections decay as $1/\ln L$, while in the infinite-volume
case, see Eq.~(\ref{R-infvol}), they decay as $1/(\ln \xi_\infty)^2$,
i.e.  with the {\em square} of the logarithm of the relevant length
scale.

The finite-size behavior of observables ${\cal O}$ with anomalous RG
dimension can be obtained in a similar fashion.  We need to compute
the behavior of the integral
\begin{eqnarray}
&& \int_{v_0}^{v(L)} dw {\gamma_{\cal O}(0,w) \over \beta_v(0,w)} = 
   \int^{v_0}_{v(L)} dw 
  {\gamma_{\cal O}(0,w) \over F(w)[1 + w f(w^2)]} = \nonumber \\
&& = 
   g_{00} \ln L + 
\int^{v_0}_{v(L)} dw
    {\gamma_{\cal O}(0,w) -g_{00} \over F(w)[1 + w f(w^2)]} =\\
&&= g_{00} \ln L - g_{01} \ln v(L) + K' + O(v),\nonumber
\end{eqnarray}
where $\gamma_{\cal O}(Q,v)$ is the anomalous dimension as a function
of $Q$ and $v$, and $K'$ is a nonuniversal constant. Hence
\begin{equation}
Z[0,v(L)] = L^{g_{00}} v(L)^{-g_{01}} e^{K'} [1 + O(v) ]
    Z(0,v_0) .
\label{Zscal-FSS}
\end{equation}
We end up with 
\begin{equation}
{{\cal O}(0,v_0,L) \over L^{d_{\cal O} + g_{00}}} = 
     K'   \left( \mu+{1\over 2}\ln\mu\right)^{g_{01}} 
    \left[1 + O(\mu^{-1})\right].
\end{equation}

The above results imply that the two-point function at $T_c$ behaves as
\begin{equation}
G({\bf x},{\bf y}) \approx  L^{-1/4}  \left( \mu+{1\over 2}\ln\mu\right)^{1/8}
 {\cal G}({\bf x}/L,{\bf y}/L).
\label{gxylofss}
\end{equation}
In particular, its space integral, i.e. the  susceptibility, scales at the critical point as 
\begin{equation}
\chi L^{-7/4} = \hat{K} \left( \mu+{1\over 2}\ln\mu\right)^{1/8} 
    \left[1 + O(\mu^{-1})\right],
\label{chi-FSS}
\end{equation}
where we used $g_{01} = 1/8$. 
A numerical analysis of the 2D XY model providing evidence of the
leading multiplicative logarithm is reported in
Refs.~\cite{Hasenbusch-05,KO-12}.
 Note that a naive integration of the
infinite-volume two-point function $G(r)$ up to $r\sim L$ would give
the same result, $\chi \sim L^{7/4} (\ln L)^{1/8}$. The critical-point
behavior (\ref{chi-FSS}) should be contrasted with Eq.~(\ref{chi-expRG}):
in infinite-volume $\chi \sim \xi_\infty^{7/4}$ without additional 
leading logarithms.

\section{Trap-size scaling}
\label{TSSsec}

Statistical systems are generally inhomogeneous in nature, while
homogeneous systems are often an ideal limit of experimental
conditions.  Thus, in the study of critical phenomena, an important
issue is how critical behaviors develop in inhomogeneous systems.
Particularly interesting physical systems are interacting particles
constrained within a limited region of space by an external force.
This is a common feature of recent experiments with diluted atomic
vapors~\cite{CWK-02} and cold atoms in optical lattices~\cite{BDZ-08},
which have provided a great opportunity to investigate the interplay
between quantum and statistical behaviors in particle systems.

Experimental evidences of BKT transitions in trapped quasi-2D atomic
gases have been reported in
Refs.~\cite{HKCBD-06,KHD-07,HKCRD-08,CRRHP-09,HZGC-10}.  The
inhomogeneity due to the trapping potential drastically changes, even
qualitatively, the general features of the critical behavior.  For
example, the correlation functions of the critical modes do not
develop a diverging length scale in a trap.  Nevertheless, when the
trap size becomes large the system develops a critical scaling
behavior, which can be described in the framework of the TSS
theory~\cite{CV-09,CV-10}.  TSS has some analogies with the standard
FSS for homogeneous systems with two main differences: the
inhomogeneity due to the space-dependence of the external field, and a
nontrivial dependence of the correlation length on the trap size at
the critical point.

The above considerations apply to general quasi-2D systems of
interacting bosonic particles trapped by an external harmonic
potential. In particular, we mention systems of bosonic cold atoms in
quasi-2D optical lattices~\cite{BDZ-08}, which can be effectively
described~\cite{JBCGZ-98} by the Bose-Hubbard (BH)
model~\cite{FWGF-89}
\begin{eqnarray}
H_{\rm BH} &=& - {J\over 2} \sum_{\langle ij\rangle} (b_i^\dagger b_j+
b_j^\dagger b_i) 
\label{bhm}\\
&&+ {U\over 2} \sum_i n_i(n_i-1) - \mu \sum_i n_i,\nonumber
\end{eqnarray}
where $b_i$ is the bosonic operator, $n_i\equiv b_i^\dagger b_i$ is
the particle density operator, and the sums run over the bonds
${\langle ij \rangle }$ and the sites $i$ of a square lattice.  The
phase diagram of the 2D BH model  presents finite-temperature BKT
transition lines, connecting $T=0$ quantum transitions, such as those
from the vacuum state to the superfluid phase, and from the superfluid
phase to Mott phases~\cite{FWGF-89}.  Experiments with cold
atoms~\cite{CWK-02,BDZ-08} are usually performed in the presence of a
trapping potential, which can be taken into account by adding a
corresponding term in the Hamiltonian:
\begin{eqnarray}
&& H_{\rm tBH} =  H_{\rm BH} + \sum_i V(r_i) n_i, 
\label{bhmt}\\
&& V(r)= w^p r^p,\label{potential}
\end{eqnarray}
where $r$ is the distance from the center of the trap, and $p$ is a
positive even exponent.  A natural definition of trap size is provided
by
\begin{equation}
l_t\equiv J^{1/p}/w.  
\label{trapsize}
\end{equation}
Far from the origin the potential $V(r)$ diverges, therefore $\langle
n_i\rangle$ vanishes and the particles are trapped.  The trapping
potential is effectively harmonic in most experiments, i.e. $p=2$.

The trapped 2D BH model has been numerically investigated in
Ref.~\cite{CR-12}, by quantum Monte Carlo simulations, showing that
the BKT critical behavior is significantly modified by the presence of
the trap. Analogously, an accurate experimental determination of the
critical parameters, such as the critical temperature, critical
exponents, etc..., in trapped particle systems requires a quantitative
analysis of the trap effects.  In the following we investigate this
issue at the BKT transition and derive the asymptotic TSS behavior
from the BKT RG flow.

\subsection{General features of trap-size scaling}
\label{genfea}

Let us first describe the general TSS approach to standard continuous
transitions~\cite{PV-r}, characterized by two relevant parameters
$\tau$ and $h$ (usually $\tau\sim T/T_c-1$ and $h$ is the external
field coupled to the order parameter), whose RG dimensions are
$y_\tau=1/\nu$ and $y_h= (d+2-\eta)/2$.  The presence of an external
space-dependent field $V(r)=(w r)^p$ significantly affects the
critical modes, introducing another length scale, the trap size
$l_t\sim 1/w$.  Within the TSS framework~\cite{CV-09,CV-10}, the
scaling of the singular part of the free-energy density around the
center of the trap is generally written as
\begin{equation}
F({\bf x},T,h) = l_t^{-\theta d}
{\cal F}(rl_t^{-\theta},\tau l_t^{\theta y_\tau},hl_t^{\theta y_h}),
\label{freee}
\end{equation}
where $r$ is the distance from the center of the trap, and $\theta$ is
the {\em trap} exponent.  TSS implies that at the critical point
($\tau=0$) the correlation length $\xi$ of the critical modes is
finite, but increases as $\xi \sim l_t^{\theta}$ with increasing the
trap size $l_t$.  TSS equations can be derived for the correlation
functions of the critical modes. For example, the correlation function
of the fundamental field $\psi(x)$ (the quantum field $b$ in the BH
model) is expected to behave as
\begin{equation}
G({\bf x},{\bf y}) \equiv \langle \bar{\psi}({\bf x}) \psi({\bf y})
\rangle_c = l_t^{-\theta\eta} {\cal G}({\bf x} l_t^{-\theta},{\bf y}
l_t^{-\theta},\tau l_t^{\theta/\nu}),
\label{twopfpl}
\end{equation}
where ${\cal G}$ is a scaling function.  

The trap exponent $\theta$ generally depends on the universality class
of the transition, on its space dependence (in experiments the
external potential is usually harmonic), and on the way it couples to
the particles.  Its value can be inferred by a RG analysis of the
perturbation $P_V$ representing the external trapping potential
coupled to the particle density.  The universality class of the
superfluid transition can be represented by a $\Phi^4$ theory for a
complex field $\psi$ associated with the order parameter, see, e.g.,
Ref.~\cite{ZJ-book},
\begin{equation}
H_{\Phi^4} = \int d^d x                                        
\left[ |\partial_\mu \psi({\bf x})|^2 + 
r |\psi({\bf x})|^2 + u |\psi({\bf x})|^4\right].
\label{hphi4}
\end{equation} 
Since the particle density corresponds to the energy operator
$|\psi|^2$, we write the perturbation $P_V$ as
\begin{equation}
P_V=\int d^d x\, V({\bf x}) |\psi({\bf x})|^2.
\label{pertu}
\end{equation}
The exponent $\theta$ is related to the RG dimension $y_w$ of the coupling $w$
of the external field $V=(w r)^p$ by $\theta=1/y_w$.  Then, a 
standard RG argument gives 
\begin{eqnarray}
py_w - p + y_n = d,\label{rg1}
\end{eqnarray} 
where $y_n=d-1/\nu$ is the RG dimension of the density/energy operator
$|\psi|^2$. We eventually obtain
\begin{equation}
\theta = {1\over y_w}= {p\nu\over 1 + p \nu}.
\label{thetap}
\end{equation}
We may derive the value of $\theta$ at the BKT transition by formally
setting $\nu=\infty$ in Eq.~(\ref{thetap}), somehow corresponding to
the BKT exponential behavior of the correlation length $\xi \sim {\rm
exp}(c\tau^{-1/2})$, where $\tau \equiv T/T_{c}-1\rightarrow 0^+$.
This would give $\theta=1$ for any power $p$ of the potential.  This
result is also obtained by extending to the BKT transition point the
result $\theta=1$ for the TSS in the whole QLRO phase~\cite{CV-11},
which can be inferred by a RG analysis of the trap perturbation along
the low-temperature line of Gaussian fixed points where spin-wave
theory applies.

The trap-size dependence predicted by TSS has been verified at various
phase transitions, for example at the Ising transition of lattice gas
models~\cite{CV-09,QSS-10}, at the 3D superfluid transition of bosonic
particle systems such as those described by the 3D BH
model~\cite{CTV-13}, and at the quantum $T=0$ Ising and Mott
transitions~\cite{CV-10,CV-10b,CTV-12,CT-12}.  Multiplicative
logarithms are generally expected at the upper dimension of the given
universality class. We shall show that they also appear at the BKT
transition, which should not be surprising because they are already
present in the scaling behavior of homogeneous systems, as discussed
in the previous sections.

Note that the RG dimension associated with the size $L$ is also 1 (in
length units). This might suggest that TSS is analogous to FSS,
i.e. characterized by the same power laws and similar multiplicative
logarithms.  However, as we shall see below, the analysis of the RG
flow taking into account the trapping potential shows that the
asymptotic trap-size dependence at the BKT transition presents
multiplicative logarithms which turn out to depend on the power law of
the trapping potential.  Therefore, at a BKT transition the TSS
relations (\ref{freee}) and (\ref{twopfpl}) must be revised, including
multiplicative logarithms, which differ from those observed in the FSS
case.

\subsection{Renormalization-group analysis of trap-size scaling}
\label{astss}

To investigate the TSS regime, we extend the RG analysis presented above.
It is quite obvious
that the presence of the trap does not change the short-distance
behavior of the model, hence no change should be made on the scaling
behavior of the couplings. Let us now consider the flow of the
coupling $w$ entering  the potential (\ref{bhmt}), which, in full
generality, can have the form
\begin{equation}
{d w \over dl} = \beta_w(u,v,w).
\end{equation}
If we start from $w = 0$ (no trap), we should always have $w(l) = 0$,
hence the $\beta$-function should have the form
\begin{equation}
\beta_w(u,v,w) = w H(u,v,w).
\label{beta-trap-ini}
\end{equation}
Assuming $y_w=\theta^{-1}=1$, we have
\begin{equation}
H(0,0,0)=y_w=1. 
\label{thetah0}
\end{equation}
In App.~\ref{app:betatrap} we show that we can define a nonlinear
scaling field $z(u,v,w)$ so that the $\beta$-function
(\ref{beta-trap-ini}) becomes
\begin{equation}
\beta_z(u,v,z) = z T(Q,v),
\label{beta-trap}
\end{equation}
where, as before, we have replaced $u$ with the RG invariant quantity
$Q$ and $T(0,0) = 1$.  The RG flow of $z$ is particularly
simple:
\begin{eqnarray}
z(l) &=& z_0 \exp\left\{ \int_0^l T[Q,v(l')] dl'\right\} 
\label{TSS-zlgen}
\\
     &=& z_0 e^{l} 
   \exp\left\{ -\int_{v_0}^{v(l)} dw\
      {T(Q,w) - 1\over [Q + F(w)][1 + w f(w^2)]} \right\},
\nonumber 
\end{eqnarray}
where $z_0\sim 1/l_t$ is the starting point of the flow.  Below, we
shall consider two cases: first, we consider the high-temperature
phase $T>T_c$, then TSS at the critical point. In both cases, we
assume that the infinite-volume limit has been attained, i.e. that
$L\gg l_t$.

\subsubsection{TSS in the high-temperature phase}
\label{astssht}

We start from the general scaling relation for the two-point 
correlation function
\begin{equation}
G({\bf x},{\bf y};Q,v_0,z_0) = 
   Z[Q,v(l)] G[{\bf x} e^{-l},{\bf y} e^{-l}, Q, v(l), z(l)].
\label{TSS-Gscal}
\end{equation}
Note that the renormalization function $Z(Q,v)$ does not depend on the
scaling field $z$, since the renormalization constant is only
determined by the short-distance behavior of the operators --- the
fundamental field in this case --- defining the correlation function.
As in Sec.~\ref{infvolsec}, we fix $l=l_0$ by requiring $v(l_0) = -1$.
Then, by using Eq.~(\ref{infvol-xiQv0}) we can write
\begin{equation} 
   e^{l_0} = {\xi_\infty(Q,v_0)\over \xi_\infty(Q,-1)} 
   \approx a \xi_\infty(Q,v_0) [1 + O(Q)],
\end{equation}
where $a = 1/\xi_\infty(Q=0,-1)$ is a constant. To obtain the trap
corrections, we must evaluate $z(l_0)$. Eq.~(\ref{TSS-zlgen}) allows
us to write
\begin{eqnarray}
z(l_0) &=& z_0 a \xi_\infty(Q,v_0)    \\
   && \times \exp\left\{ -\int_{v_0}^{-1} dw\
      {T(Q,w) - 1\over [Q + F(w)][1 + w f(w^2)]} \right\}.
\nonumber 
\end{eqnarray}
The integral is finite for $Q\to 0$, with corrections of order
$\sqrt{Q}$, see App.~\ref{App.B}.  Hence
\begin{equation}
z(l_0) \sim z_0 \xi_\infty(Q,v_0) [1 + O(Q^{1/2})].
\end{equation}
Since  $l_t \sim 1/z_0$, 
 we obtain the general scaling form
\begin{equation}
G({\bf x},{\bf y};\tau) = 
   {\cal G}\left(
     {\bf x}\xi^{-1}_\infty, {\bf y}\xi^{-1}_\infty,
     l_t\xi^{-1}_\infty\right) + O(\sqrt{Q}).
\end{equation}

\subsubsection{TSS at the critical point}
\label{astssTC}

Let us now consider TSS at criticality ($Q = 0$). In this case it is 
convenient to fix $l=l_0$ by setting $z(l_0) = 1$. 
For $l\to \infty$, $v(l) \to 0$, hence we can rewrite Eq.~(\ref{TSS-zlgen}) as
\begin{equation}
1 = z_0 T_0 e^{-l_0} v(l_0)^{-t_1} [1 + O(v(l_0))] ,
\end{equation}
where $t_1$ is defined by the expansion 
\begin{equation}
T(0,w) = 1 + t_1 w + O(w^2),\label{t0wexp}
\end{equation}
and  $T_0$ is a nonuniversal constant given by 
\begin{equation}
T_0 = v_0^{-t_1} 
   \exp\left\{ \int_{0}^{v_0} dw\
      \left[{T(0,w) - 1\over F(w)[1 + w f(w^2)]} - {t_1\over w}\right]\right\}.
\end{equation}
The flow of $v(l)$ follows from Eq.~(\ref{flow-vl-basic}).  Using the
results of Sec.~\ref{cpfss} and in particular Eq.~(\ref{lnL_flow}), we
have $v(l)=l^{-1} + O(l^{-2}\ln l)$, so we obtain
\begin{equation}
1 = z_0 T_0 e^{l} l^{t_1} [1 + O(l^{-1}\ln l)] .
\end{equation}
Inverting this equation we obtain
\begin{eqnarray}
e^{l} \approx {|\ln z_0 T_0|^{-t_1} \over z_0 T_0}.
%\qquad  l \approx \ln z_0 T_0 ( 1 + O (\ln\ln z_0 T_0/\ln z_0 T_0) ).
\end{eqnarray}
To obtain the scaling behavior, we need to determine the large-$l$
behavior of $Z[0,v(l)]$.  Using the results of Sec.~\ref{cpfss} and in
particular Eq.~(\ref{Zscal-FSS}), we obtain
\begin{eqnarray}
Z[0,v(l)] &\approx & e^{l g_{00}} v(l)^{-g_{01}} e^{K'} Z(0,v_0) \nonumber \\
&\approx&  {|\ln z_0 T_0|^{-t_1 g_{00} + g_{01}} \over 
            (z_0 T_0)^{g_{00}} } e^{K'} Z(0,v_0).
\end{eqnarray}
Substituting this result into Eq.~(\ref{TSS-Gscal}), and choosing the
length scale so that $z_0 T_0 \approx 1/l_t + O(l_t^{-2})$, we obtain
the TSS of the two-point function at $T_c$.  We write it as (using
$g_{00} = -1/4$ and $g_{01} = 1/8$)
\begin{eqnarray}
G({\bf x},{\bf y}) &=& l_t^{-1/4} (\ln l_t)^{1/8 + \kappa/4} \times
\label{gxylo}\\
& \times &  {\cal G}[{\bf x}(\ln l_t)^{\kappa}/l_t,
        {\bf y}(\ln l_t)^{\kappa}/l_t ]\nonumber,
\end{eqnarray}
where $\kappa=t_1$.

We may also define the trap susceptibility $\chi_t$, 
\begin{equation}
\chi_t = \sum_{\bf x} G({\bf 0},{\bf x}), \label{defchitss}
\end{equation}
and the trap correlation length $\xi_t$,
\begin{equation}
\xi_t^2 = {1\over 4 \chi_t} \sum_{\bf x} |{\bf x}|^2 G(0,{\bf x}) .
\label{defxitss}
\end{equation}
Eq.~(\ref{gxylo}) implies the asymptotic behaviors
\begin{eqnarray}
&&\chi_t  \sim  l_t^{7/4} (\ln l_t)^{1/8-7\kappa/4},\label{chitss}\\
&&\xi_t \sim  l_t (\ln l_t)^{-\kappa}.  \label{xitss} 
\end{eqnarray}
We do not know the value of the coefficient $t_1$ appearing in the
expansion (\ref{t0wexp}), which provides the exponent $\kappa$ in the
asymptotic formulas.  We generally expect that it depends on the power
$p$ of the trapping potential. Note that $\kappa$ must vanish in the
limit $p\to\infty$. Indeed, in this limit the trapped system is
equivalent to a homogeneous system confined in a circle of radius $l_t
= 1/w$ with open boundary conditions.  Therefore, TSS is equivalent to
standard FSS with $L\sim l_t$, hence $\kappa = 0$.  We anticipate that
the numerical analysis that we present below provides a strong
evidence that $\kappa$ depends on $p$; in particular, it suggests
$\kappa=2/p$.

\subsection{Numerical results in the presence of an external
space-dependent field}
\label{numtest}

The main features of TSS are expected to be universal, hence the RG results 
should apply to generic 2D systems characterized by a U(1)
symmetry in the presence of an
external space-dependent field coupled to the energy density.  For a
numerical check of the RG predictions, we consider the classical 2D
XY model in the presence of 
an external space-dependent field coupled to the energy density.
The Hamiltonian is given by
\begin{eqnarray}
&& H_U = -  J \sum_{\langle ij \rangle }
 {\rm Re} \, \bar{\psi}_i U_{ij} \psi_j,
\label{xymodtr} \\
&& U_{ij} = [1 + V(r_{ij})]^{-1}, \quad V(r)=w^pr^p,
\label{hoppingt}
\end{eqnarray}
where $p$ is an even positive integer, $r_{ij}$ is the distance from
the origin of the midpoint of the lattice link connecting the
nearest-neighbor sites $i$ and $j$. We set $J=1$.  The inhomogeneity
arising from the space dependence of $U_{ij}$ is analogous to that
arising from a trapping potential in particle systems.  Thus, $l_t
\equiv 1/w$ may be considered as the analog of the trap size
(\ref{trapsize}).  For $p\to \infty$, $V(r) = 0$ for $r < l_t$ and
$V(r) = \infty$ for $r > l_t$. Hence, the system is equivalent to a
homogeneous system confined in a circle of radius $l_t$ with open
boundary conditions.  Therefore, in this limit TSS must reproduce
standard FSS.  A study of the trap effects in the low-temperature
phase is reported in Ref.~\cite{CV-11}.  Here we focus on the
trap-size dependence at the BKT critical temperature
$T_c=0.89294(8)$~\cite{HP-97,Hasenbusch-05} of the homogeneous XY
model (\ref{XYmodel}), which is also the model (\ref{xymodtr}) with
$U_{ij}=1$.

\begin{figure}[tbp]
\includegraphics*[scale=\graphicscale]{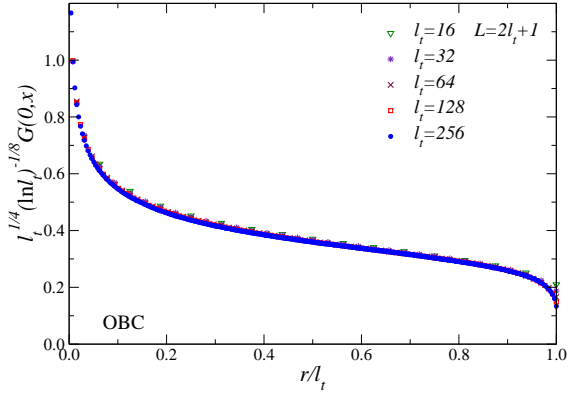}                               
\caption{FSS plot of the two-point function $G({\bf 0},{\bf x})$ at $T_c$
   for ${\bf x} = (r,0)$, $0\le r \le l_t=(L-1)/2$,
  for a homogeneous XY model on a $L^2$ lattice with open boundary
  conditions (OBC).}
\label{gpinf}
\end{figure}

We present results of Monte Carlo (MC) simulations of model
(\ref{xymodtr}). We use a mixture of Metropolis and overrelaxation
updates of the spin variables~\cite{HPV-05}.  We consider square
lattices with $L^2$, odd $L$, sites and open boundary
conditions. Lattice points have coordinates $(x,y)$ with $-(L-1)/2\le
x,y\le (L-1)/2$, so that the origin $(0,0)$ is at the center of the
lattice.  The external potential is given by $V({\bf x})=(r/l_t)^p$,
where $r\equiv |{\bf x}|$ and $l_t$ is the trap size.  The lattice
size $L$ is taken sufficiently large to avoid finite-size effects. We
check that they are negligible compared with the statistical errors by
comparing results at fixed trap size $l_t$ for different lattice sizes
$L$.

\begin{figure}[tbp]
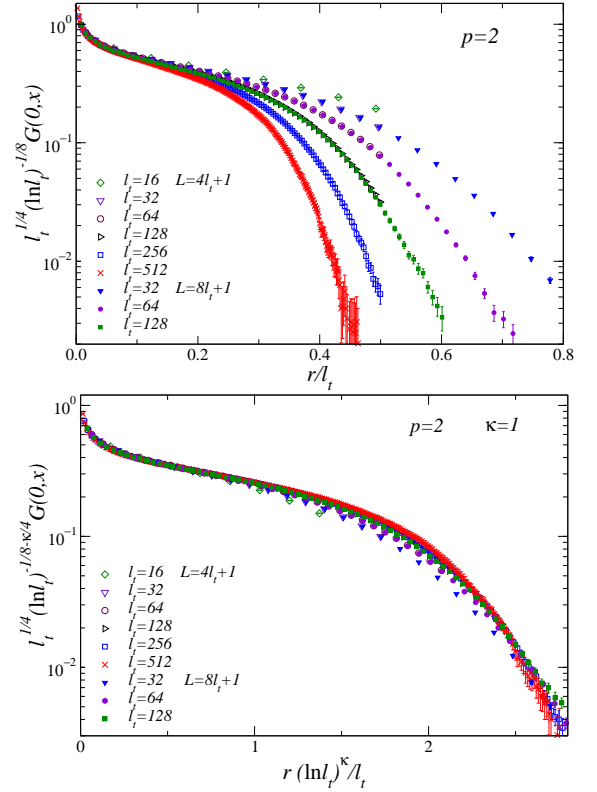

\includegraphics*[scale=\graphicscale]{fig3a.eps}
\includegraphics*[scale=\graphicscale]{fig3b.eps}
\caption{TSS plot of the two-point function $G({\bf 0},{\bf x})$ at
  $T_c$ for a harmonic potential $V=(r/l_t)^2$.  We set $\kappa=0$
  (top) and $\kappa=1$ (bottom).  To check finite-size effects, we
  report two data sets: $L\approx 4l_t$ and $L\approx 8l_t$ in the two
  cases, respectively.}
 \label{gp2}
\end{figure}

We want to check the TSS equation (\ref{gxylo}) for the correlation
function at $T_c$.  For this purpose we report results at $T_c$ for
the correlation function $G({\bf 0},{\bf x}) \equiv \langle
\bar{\psi}({\bf 0}) \psi({\bf x})\rangle$, which is expected to scale
as
\begin{equation}
G({\bf 0},{\bf x}) =
l_t^{-1/4} (\ln l_t)^{1/8+\kappa/4}\,
 {\cal G}_p\left[ r(\ln l_t)^{\kappa}/l_t\right],
\label{g0xp}
\end{equation}
where $r\equiv |{\bf x}|$, and the exponent $\kappa$ is expected to
depend on the power of the external potential.

\begin{figure}[tbp]
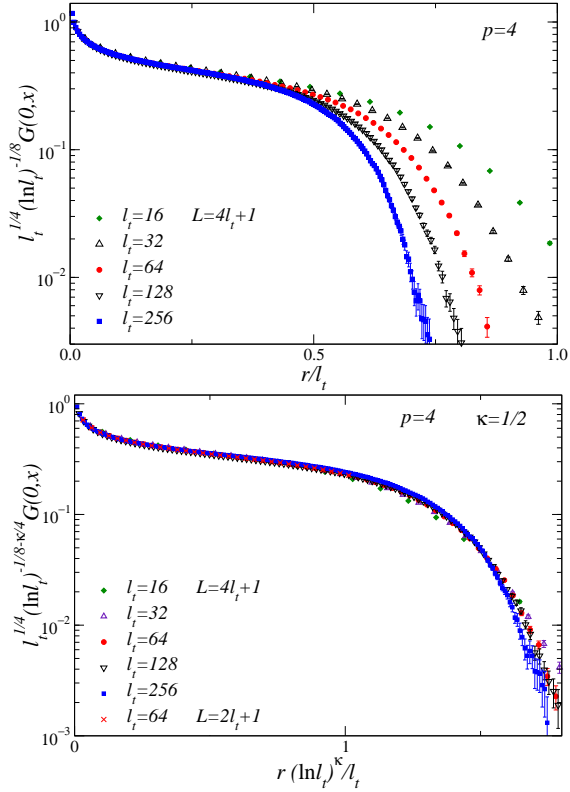

\includegraphics*[scale=\graphicscale]{fig4a.eps}
\includegraphics*[scale=\graphicscale]{fig4b.eps}
\caption{TSS plot of the two-point function $G({\bf 0},{\bf x})$ at
  $T_c$ for a trap with potential $V=(r/l_t)^4$.  We set $\kappa=0$
  (top) and $\kappa=1/2$ (bottom).  To check finite-size effects, we
  report data for $L\approx 4l_t$ and $L\approx 2l_t$.}
 \label{gp4}
\end{figure}

We first check the FSS behavior of the homogeneous XY model with open
boundary conditions (OBC), which is formally equivalent to the limit
$p\to\infty$ of model (\ref{hoppingt}) with $L=2l_t+1$ and $l_t$
integer. Note that translation invariance is lost in systems with
OBC. Fig.~\ref{gpinf} shows the data for ${\bf x}=(r,0)$, which
clearly support the expected FSS behavior
\begin{equation}
G({\bf 0},{\bf x}) = L^{-1/4} (\ln L)^{1/8} g(x/L).
\label{g0xpi}
\end{equation}
Note that OBC breaks translation invariance, and gives rise to
power-law boundary corrections~\cite{Diehl-86,Hasenbusch-12} which are
expected to be $O(L^{-1})$.  In Figs.~\ref{gp2} and \ref{gp4} we show
the results of the simulations for two different values of $p$, $p=2$
and $p=4$.  In order to check the scaling behavior (\ref{g0xp}), we
present TSS plots with $\kappa=0$, using the analog of the FSS formula
(\ref{g0xpi}), and with an optimal nonzero value of the coefficient
$\kappa$, which is determined by looking for the best collapse of the
data. Optimal scaling is obtained be setting $\kappa\approx 1$ for
$p=2$ and $\kappa\approx 1/2$ for $p=4$, with an uncertainty which we
estimate to be approximately 10\%.  This simple scaling test clearly
favors a nonzero $p$-dependent value for $\kappa$.  Taking also into
account that $\kappa=0$ for $p\to\infty$, the above numerical results
hint at the simple formula $\kappa=2/p$.  These results should be
universal, hence they also apply to the BKT transitions of general
systems of 2D interacting bosonic particles trapped by an external
space-dependent potential, such as those which have been investigated
experimentally~\cite{HKCBD-06,KHD-07,HKCRD-08,CRRHP-09,HZGC-10,BDZ-08}.

\begin{figure}[tbp]
\includegraphics*[scale=\graphicscale]{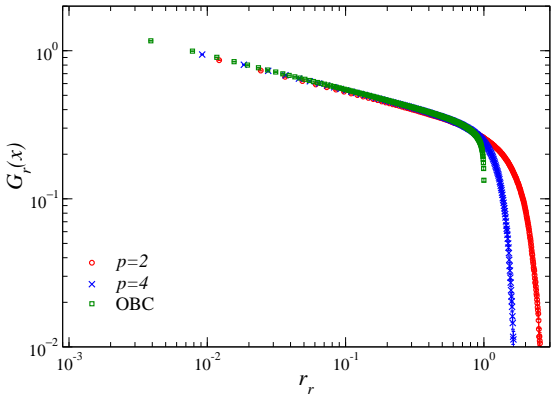}
\caption{Plot of
$G_r({\bf x})\equiv l_t^{1/4} (\ln
l_t)^{-1/8-\kappa/4} G({\bf 0},{\bf x})$ vs. $r_r\equiv r(\ln
l_t)^{\kappa}/l_t$ for $p=2$, $p=4$, and for the homogeneous
system with OBC (it corresponds to
$p\to\infty$), using the data for the largest available trap sizes.  We
use $\kappa=1,\,1/2,\,0$ for $p=2,\,4$ and the OBC case, respectively.  }
 \label{gpall}
\end{figure}

Finally, in Fig.~\ref{gpall} we compare the scaling functions
associated with $G({\bf 0},{\bf x})$ for $p=2$, $p=4$, and for the
homogeneous system with OBC. We plot $G_r({\bf x})\equiv l_t^{1/4}
(\ln l_t)^{-1/8-\kappa/4} G({\bf 0},{\bf x})$ versus the scaling
variable $r_r\equiv r(\ln l_t)^{\kappa}/l_t$ for the largest available
trap size or lattice, using $\kappa=2/p$.  For small $r_r$ all curves
run very close up to $r_r$ of order one. This behavior is not
surprising, since for $r/l_t\to 0$ boundary effects should become
irrelevant, hence all scaling functions should behave in the same
manner.

\section{Conclusions and summary of the main results}
\label{secco}

We have investigated the general features of the RG flow at the BKT
transition, providing a definite characterization of the asymptotic
critical behavior including the universal multiplicative or subleading
logarithmic corrections, in different critical regimes: (i) in the
infinite-volume critical disordered phase; (ii) in the FSS limit, both
for $T>T_c$ and $T=T_c$; (iii) in the TSS limit, again both above and
at the critical temperature.

For this purpose, we exploit the mapping between the standard XY or
Coulomb gas models and the SG model, whose RG flow around the
renormalizable fixed point describes the BKT transition~\cite{AGG-80}.
To determine the RG flow, we first derive a simple canonical universal
expression for the $\beta$-functions.  Then, we determine the
asymptotic solutions of the RG equations in the different situations
mentioned above.  For the TSS case, numerical simulations confirm the
RG predictions.

We summarize our main results:

(a) We prove that an appropriate analytical redefinition of the
couplings of the SG model, cf. Eq.~(\ref{sineG}), allows us to write
the $\beta$-functions as
\begin{eqnarray}
&&\beta_u(u,v) = - u v, \label{betauvco}\\
&&\beta_v(u,v) = -
u^2 [1 + v f(v^2)],
\nonumber
\end{eqnarray}
to all orders of perturbation theory. The universal function $f(x)$
can be expanded as $f(x) = b_0 + b_1 x + b_2 x^2 + ...$. The
zeroth-order term $b_0=3/2$ can be obtained from the two-loop results
of Refs.~\cite{AGG-80,BH-00}, while the next coefficient $b_1$ can
only be determined by means of a four-loop calculation in the SG
model.  In practice, our results extend the knowledge of the RG flow
of the SG model to three loops.  Under the same redefinition of the
couplings, the anomalous dimension $\gamma(u,v)$ of the fundamental
field becomes $\gamma=-1/4 + v/8 - v^2/16 + g_{20} u^2 + \dots$, where
$g_{20}$ is a nonuniversal constant which does not enter the universal
scaling behavior of the two-point function.

(b) In the high-temperature critical regime and in the infinite-volume
limit, the RG flow implies the BKT asymptotic behavior (setting
$\tau=T/T_c-1$) \cite{KT-73,B-72,Kosterlitz-74,JKKN-77}
\begin{equation}
\xi_\infty(\tau) = X e^{c/\sqrt{\tau}} \left[ 1 + O(\sqrt{\tau})\right],
\label{xiKTco}
\end{equation}
for the infinite-volume 
correlation length, where $X$ is a nonuniversal constant.  
In the case of RG
invariant quantities $R$, such as the ratio of two different
definitions of correlation lengths and the zero-momentum
four-point renormalized coupling~\cite{PV-r}, we have
\begin{equation}
R(\tau) = R^* + O(1/\ln^2\xi_\infty).
\label{rtauco}
\end{equation}
Corrections of order $1/\ln\xi_\infty$ are absent.  The
susceptibility, defined as the space-integral of the two-point
function of the fundamental field, behaves as
\begin{equation}
\chi = A_\chi \xi^{7/4}_\infty 
\left[ 1 + {\pi^2\over 16 \ln(\xi_\infty/X)} + 
   O\left({1\over \ln^2\xi_\infty}\right)\right]
\label{chiasco}
\end{equation}
where $A_\chi$ is a nonuniversal amplitude. The correction term
behaving as $1/\ln \xi_\infty$ is universal. 

(c) We have studied the FSS behavior.  In the high-temperature phase,
for any observable ${\cal O}$, the ratio ${\cal O}(L)/{\cal O}(L\to
\infty)$ approaches a universal function $A(L/\xi_\infty)$ with
$O(1/\ln^2L)$ corrections.  The approach to the $L\to\infty$ limit is
not uniform in $T$. At $T=T_c$ further logarithms appear.  Indeed,
setting $ \mu \equiv \ln(L/\lambda)$ where $\lambda$ is an appropriate
nonuniversal length scale, we have
\begin{eqnarray}
R_\xi\equiv \xi/L = R_\xi^*  +  {C_\xi\over \mu  + {1\over 2} \ln\mu}
+ O(\mu^{-2}),
\label{fsstcxico}
\end{eqnarray}
where $R_\xi^*$ and $C_\xi$ are universal, depending only on the shape
of the systems and their boundary conditions.  An analogous formula is
obtained for any RG invariant quantity, such as the helicity modulus
and the Binder cumulants. Moreover, the asymptotic behavior of the
susceptibility reads
\begin{eqnarray}
\chi L^{-7/4} = \hat{A}_\chi  \left( \mu+{1\over 2}\ln\mu\right)^{1/8} 
\left[ 1 + O(\mu^{-1})\right],
\label{fsstcchico}
\end{eqnarray}
where $\hat{A}_\chi$ is a nonuniversal constant. 
Note two important differences between the behavior at $T_c$ and 
in the high-temperature critical regime (b). For RG invariant quantities 
corrections decay as $1/\ln L$ in the first case; 
in the second one corrections of order $1/\ln\xi_\infty$ are instead absent.
Second, the behavior of the susceptibility is characterized by a 
leading logarithmic term at $T_c$, while no such term is present in 
the high-temperature case, see Eq.~(\ref{chiasco}).

(d) Finally, we consider BKT transitions in 2D interacting bosonic
particles which are trapped within a limited region of space by an
external space-dependent force, which is a common feature of
experiments with diluted atomic vapors~\cite{CWK-02} and cold atoms in
optical lattices~\cite{BDZ-08}.  We investigate how the BKT critical
behavior is affected by the presence of the external space-dependent
trapping potential $V(r)= (r/l_t)^p$ coupled to the particle (energy)
density.  We consider observables derived from the two-point function
$G({\bf x},{\bf y})$ of the fundamental field describing the critical
modes, such as the one-particle correlation function $\langle
b_x^\dagger b_y\rangle$ of the BH model (\ref{bhmt}), which describes
trapped bosonic atoms in an optical lattice.  The analysis of the RG
flow shows that TSS at the BKT transitions is characterized by a trap
exponent $\theta=1$, with additional multiplicative logarithms at
$T=T_c$.  For example, at $T_c$ the two-point function scales as
\begin{equation}
G({\bf 0},{\bf x}) = l_t^{-1/4} (\ln l_t)^{1/8+\kappa/4}\,
 {\cal G}_p\left[r(\ln l_t)^\kappa/l_t\right],
\label{g0xpc}
\end{equation}
where $\kappa$ is a new exponent which arises from the analysis of the
RG flow of the external potential and which is expected to depend on
the power $p$ characterizing the trap potential. 
Of course, in the limit $p\to\infty$ we
must have $\kappa\to 0$, since we must recover the known FSS behavior of a 
homogeneous systems.  
The scaling equation (\ref{g0xpc}) implies also
that the correlation length $\xi_t$ of the critical modes behaves
asymptotically as
\begin{eqnarray}
\xi_t \sim  l_t (\ln l_t)^{-\kappa}. \label{xitssc} 
\end{eqnarray}
These results are supported by 
Monte Carlo simulations of a 2D XY model with a space-dependent
potential coupled to the energy density.
They provide a clear evidence of the multiplicative
logarithms in Eqs.~(\ref{g0xpc}) and (\ref{xitssc}) and are 
numerically consistent with the conjecture $\kappa=2/p$.

\bigskip

{\em Acknowledegment}: We thank Gianpaolo Cicogna and Martin
Hasenbusch for useful discussions and correspondence.

\appendix 

\section{Canonical form of the BKT $\beta$-functions }
 \label{app:beta}

In the following we prove that the SG $\beta$-fuctions can be 
simplified by a redefinition of the couplings, reducing them to the
canonical form given by Eqs.~(\ref{betau}) and (\ref{betav-1}).  For a
general discussion of the mathematical problem of the reduction of
coupled differential equations to canonical form, see, e.g.,
Refs.~\cite{CG-99,Arnold}.

To all orders in the couplings $\alpha$ and $\delta$,
the $\beta$-functions of the SG model have the generic form
\begin{eqnarray}
&&\beta_\alpha = -2\alpha\delta +
   \sum_{n+m>2} b_{\alpha,nm} \alpha^n \delta^m, \\
&& \beta_\delta = -{1\over 32} \alpha^2 +
   \sum_{n+m>2} b_{\delta,nm} \alpha^n \delta^m.
\end{eqnarray}
In the SG model the sign of $\alpha$ is irrelevant, which implies the
symmetry relations
\begin{eqnarray}
&&\beta_\alpha(\alpha,\delta) = - \beta_\alpha(-\alpha,\delta) ,\\
&&\beta_\delta(\alpha,\delta) =  \beta_\delta(-\alpha,\delta). 
\end{eqnarray}
As a consequence, $b_{\alpha,nm} = 0$ if $n$ is even and
$b_{\delta,nm} = 0$ if $n$ is odd. Moreover, for $\alpha = 0$ the
theory is free and $\delta$ does not flow. Hence
\begin{equation}
\beta_\delta(\alpha=0,\delta)  = 0,
\end{equation}
which implies $b_{\delta,nm} = 0$ if $n=0$.

We wish now to prove that, by an analytic redefinition of the
couplings,
\begin{eqnarray}
\alpha &=& a_{\alpha,10} u +
   \sum_{n+m\ge 2} a_{\alpha,nm} u^n v^m, \\
\delta &=& a_{\delta,10} v +
   \sum_{n+m\ge 2} a_{\delta,nm} u^n v^m,
\end{eqnarray}
we can rewrite the $\beta$-functions 
of the SG model in the form
\begin{eqnarray}
\beta_u(u,v) &=& - u v, \\
\beta_v(u,v) &=& - u^2 \left(1 + \sum_{k=0}^\infty b_{k} v^{2k+1}\right).
\nonumber
\end{eqnarray}

To prove the general result, we shall work by perturbative induction.
We assume that we have already proved the result to order $n-1$, i.e.
that we redefined couplings so that
\begin{eqnarray}
\beta_u &=& - u v + u \sum_{k=n}^\infty  H_{k-1}(u,v), \\
\beta_v &=& - u^2 \left(1 + \sum_{k=0}^{m-2} b_{k} v^{2k+1}\right) 
    + u^2 \sum_{k=n}^\infty G_{k-2}(u,v),\nonumber
\end{eqnarray}
where $m = \lfloor n/2\rfloor$, and 
$H_k(u,v)$ and $G_k(u,v)$ are homogeneous polynomials satisfying
\begin{eqnarray}
&& H_k(\lambda u,\lambda v) = \lambda^{k} H_k(u,v), \nonumber \\
&& G_k(\lambda u,\lambda v) = \lambda^{k} G_k(u,v).
\end{eqnarray}
Moreover, they are even functions of $u$. 
We now consider the change of variables
\begin{eqnarray}
&&   u' = u + u A_{n-2}(u,v),\\ 
  && v' = v + B_{n-1}(u,v),\nonumber
\end{eqnarray}
where $A_{n-2}(u,v)$ and $B_{n-1}(u,v)$ are homogeneous polynomials even
in $u$ and satisfying
\begin{eqnarray}
&& A_{n-2}(\lambda u,\lambda v) = \lambda^{n-2} A_{n-2}(u,v), \nonumber \\
&& B_{n-1}(\lambda u,\lambda v) = \lambda^{n-1} B_{n-1}(u,v).
\end{eqnarray}
We wish to show that, by a proper choice of $A_{n-2}$ and $B_{n-1}$,
we can cancel all terms of order $n$ except, if $n$ is odd,
the term
of order $u^2 v^{n-2}$ in $\beta_v$. Hence we can obtain
\begin{eqnarray}
\beta_{u'}(u',v') &=& - u' v' + 
   u' \sum_{k=n+1}^\infty  \tilde{H}_{k-1}(u',v'),\quad \\
\beta_{v'}(u',v') &=& - {u'}^2 
    \left(1 + \sum_{k=0}^{m'-2} b_{k} {v'}^{2k+1}\right) \nonumber \\
& +& {u'}^2 \sum_{k=n+1}^\infty \tilde{G}_{k-2}(u',v'),
\nonumber
\end{eqnarray}
with different $\tilde{H}$ and $\tilde{G}$ and $m' =\lfloor
(n+1)/2\rfloor$.

Requiring all terms (except the one mentioned above) to cancel, we
obtain the equations
\begin{eqnarray}
&& u v A_{n-2} + u B_{n-1} - u v {\partial (u A_{n-2})\over \partial u} 
\label{qq1}\\
&&\qquad    -  u^3 {\partial A_{n-2}\over \partial v} + u H_{n-1} = 0 
\nonumber
\end{eqnarray}
and
\begin{eqnarray}
&&2 u^2 A_{n-2} - u v {\partial B_{n-1}\over \partial u} - 
 u^2 {\partial B_{n-1}\over \partial v} +
\label{qq2}\\
&&\quad + u^2 G_{n-2} = R_n
\nonumber
\end{eqnarray}
where $R_n = 0$ for $n$ even, and $R_n = b_{(n-3)/2} u^2 v^{n-2}$ if $n$ is odd.
Eq.~(\ref{qq1}) gives us the function $B_{n-1}$:
\begin{equation}
B_{n-1} = u v {\partial A_{n-2}\over \partial u} +
      u^2 {\partial A_{n-2}\over \partial v} - H_{n-1}.
\end{equation}
Substituting it in Eq.~(\ref{qq2}), we obtain 
\begin{eqnarray} 
&&  u^2 v^2 {\partial^2 A_{n-2}\over \partial u^2} + 2 u^3 v 
   {\partial^2 A_{n-2}\over \partial u\partial v} + u^4 
   {\partial^2 A_{n-2}\over \partial v^2} + 
\nonumber\\
 && + u (u^2 + v^2) {\partial A_{n-2}\over \partial u} + 2 u^2 v 
   {\partial A_{n-2}\over \partial v} - 2 u^2 A_{n-2} - R_n
\nonumber \\
&& 
= u^2 G_{n-2} + u v {\partial H_{n-1}\over \partial u} + 
   u^2 {\partial H_{n-1}\over \partial v}.
\label{eq-A}
\end{eqnarray}
Now, we expand
\begin{equation}
A_{n-2}(u,v) = \sum_{k=0}^{n-2} a_k u^k v^{n-2-k}
\end{equation}
with $a_k = 0$ for $k$ odd.
Then, we must show that we can find $a_0$, $a_2$, ..., so that 
the following equations are satisfied. 
For $k$ even, with $k\ge 4$ and $k < n-1$, we must satisfy
\begin{eqnarray}
&&E_k = - k^2 a_k + (4 - 3 k + 2 k^2  + 2 n - 2 k n) a_{k-2} 
\nonumber\\
&&-(2 + k^2  + 3 n + n^2  - k (3 + 2 n)) a_{k-4} - \tilde{g}_k = 0.
\nonumber 
\end{eqnarray}
Here $\tilde{g}_k$ is the coefficient of order $u^k v^{n-k}$
in the expansion of the r.h.s. of Eq.~(\ref{eq-A}).
Moreover, we should satisfy
\begin{eqnarray}
E_n = (4-n)a_{n-2} - 2 a_{n-4} - \tilde{g}_n = 0 
   && \; \hbox{$n$ even},\nonumber
\\
E_{n-1} = 3 (3-n) a_{n-3} - 6 a_{n-5} - \tilde{g}_{n-1} = 0
   && \; \hbox{$n$ odd},\nonumber
\\
 E_2 = 2 (3-n) a_0 - 4 a_2 - \tilde{g}_2 = 0
   && \; \hbox{$n$ even}. \nonumber
\label{EQ2-1}
\end{eqnarray}
$E_2$ cannot be satisfied for $n$ odd---hence the necessity for the
coefficients $b_{k}$, which are defined by
\begin{equation}
b_{(n-3)/2} = E_2 = 2 (3-n) a_0 - 4 a_2 - \tilde{g}_2.
\end{equation}
We now redefine $a_{2k}$ for $k \ge 1$ as
\begin{equation}
a_{2k} = {(-1)^k a_0 \over k!} \prod_{j=1}^k \left({n-1\over2} - j\right)
    + c_{2k}.
\end{equation}
With this redefinition, we obtain 
\begin{eqnarray}
&& E_4 = 6(4 - n) c_2 - 16 c_4 - \tilde{g}_4 = 0  \\
&& E_k = - k^2 c_k + (4 - 3 k + 2 k^2  + 2 n - 2 k n) c_{k-2} 
\nonumber \\
&& \qquad -(2 + k^2  + 3 n + n^2  - k (3 + 2 n)) c_{k-4} - \tilde{g}_k = 0,
\nonumber
\end{eqnarray}
where $6\le k < n-1$. For $n$ even we should also consider 
\begin{eqnarray}
&& E_2 = - 4 c_2 - \tilde{g}_2 = 0,  \\
&& E_n = p_n a_0 + (4-n)c_{n-2} - 2 c_{n-4} - \tilde{g}_n = 0,
\nonumber
\end{eqnarray}
where 
\begin{equation}
p_n = {2 (-1)^{n/2}\over \sqrt{\pi}} {n\over (2-n) (n/2-2)!} 
    \Gamma(n/2-1/2),
\end{equation}
while for $n$ odd we also have
\begin{equation}
E_{n-1} = 3 (3-n) c_{n-3} - 6 c_{n-5} - \tilde{g}_{n-1} = 0.
\end{equation}
The parameter $b_{(n-3)/2}$ is defined by ($n$ odd)
\begin{equation}
b_{(n-3)/2} = E_2 = - 4 c_2 - \tilde{g}_2.  \\
\end{equation}
For $n$ even it is evident that all equations can be solved. 
$E_n$ can be satisfied by fixing $a_0$, while $E_k$, $k\le n-2$,
can be satisfied by fixing $c_k$. For $n$ odd, one parameter, $a_0$, is no 
longer present: this explains why we are not able to satisfy all equations
and we need to introduce the parameter $b_{(n-3)/2}$.
In practice, we can satisfy $E_4$ by fixing $c_4$ as a function of 
$c_2$, $E_6$ by fixing $c_6$ and so on. This 
allows us to solve all equations except $E_n$. 
However, there is still one free parameter, $c_2$. Substituting the
expressions of $c_{n-2}$ and $c_{n-4}$ as a function of $c_2$,
we obtain $E_n = \alpha_1 c_2 + \alpha_2$, with
\begin{equation}
\alpha_1 = {(-1)^{(n-3)/2} 2^{(7-n)/2} (n-2)!! \over (n/2-3/2)!}.
\end{equation}
Since $\alpha_1 \not=0$, also $E_n$ can be satisfied,
concluding the proof.

\section{Asymptotic expansions} \label{App.B}

We wish now to discuss the computation of the asymptotic behavior of 
integrals of the form 
\begin{equation}
   I = \int_a^b dw\, {h(w)\over Q + F(w)},
\end{equation}
where $b > 0$, $a < 0$, $h(w)$ and $F(w)$ are analytic functions 
and $F(w) \approx w^2$ for $w \to 0$. The integral $I$ diverges as $Q\to 0$
if $h(0)\not=0$. The leading behavior can be computed by approximating 
$F(w)\approx w^2$:
\begin{equation}
  I \approx \int_a^b dw\, {h(0)\over Q + w^2} 
    \approx {\pi h(0)\over \sqrt{Q}}.
\end{equation}
To compute the next nonanalytic term in the expansion, we consider 
\begin{equation}
   J = \int_a^b dw\, {h(w)\over [Q + F(w)]^2}.
\end{equation}
If $x = w/\sqrt{Q}$, we consider the expansion in powers of $\sqrt{Q}$ at fixed
$x$: 
\begin{equation}
{h(x\sqrt{Q})\over [Q + F(x\sqrt{Q})]^2} = 
    \sum_{n=-4} Q^{n/2} g_n(x).
\label{exphoverF}
\end{equation}
We define 
\begin{equation}
   G(x,Q) = \sum_{n=-4}^{-1} Q^{n/2} g_n(x),
\end{equation}
i.e. the sum of the terms that diverge as $Q\to 0$, and 
\begin{equation}
   g(w) = \lim_{Q\to 0} G(w/\sqrt{Q},Q)
\end{equation}
where the limit is taken at fixed $w$. It is easy to convince oneself that 
$g(w)$ gives the principal part of the Laurent series of $h(w)/F(w)^2$, 
so that $h(w)/F(w)^2 - g(w)$ is finite for $w \to 0$. We can thus rewrite 
\begin{eqnarray}
J &=& \int_a^b dw\,\left\{ {h(w)\over [Q + F(w)]^2} - 
         G(w/\sqrt{Q},Q)\right\} \nonumber\\
&+& 
    \int_a^b dw\, G(w/\sqrt{Q},Q).
\end{eqnarray}
The first integral is finite as $Q\to 0$, hence it does not contribute to the 
singular part of $J$. We can thus limit ourselves to considering the second 
term which can be rewritten as 
\begin{eqnarray}
J &\approx& \int_{a/\sqrt{Q}}^{b/\sqrt{Q}} dx 
   \Bigl[ Q^{-3/2} g_{-4}(x) + 
          Q^{-1} g_{-3}(x) +\nonumber \\
&+&          Q^{-1/2} g_{-2} (x) + 
          g_{-1}(x) \Bigr] .
\end{eqnarray}
For $x\to \infty$, we have $g_{-n}(x) \sim x^{-n}$ (it is easy to prove
it, using the fact that the expansion (\ref{exphoverF})
is indeed in powers of $x\sqrt{Q}$). This implies that 
\begin{equation}
\int_{a/\sqrt{Q}}^{b/\sqrt{Q}} dx\, g_{-n}(x) \approx 
   \int_{-\infty}^\infty dx\, g_{-n}(x) + O(Q^{(n-1)/2})
\end{equation}
for $n=2,3,4$. For $n=1$ we must be a little more careful. Assume that 
$g_{-1}(x) \approx g_{-1,\infty}/x$ for $x\to \infty$. Then,
\begin{eqnarray}
&&\int_{a/\sqrt{Q}}^{b/\sqrt{Q}} dx\, 
  g_{-1}(x) = \\ 
&&  = \int_{a/\sqrt{Q}}^{b/\sqrt{Q}} dx\, 
   \left[g_{-1}(x) - {g_{-1,\infty} x\over 1 + x^2}\right]
  + \int_{a/\sqrt{Q}}^{b/\sqrt{Q}} dx\, 
    {g_{-1,\infty} x\over 1 + x^2} 
\nonumber
\end{eqnarray}
The first integral now decays as $1/x^2$, hence we can extend the 
integration limits to $\pm \infty$ with corrections of order $\sqrt{Q}$; 
the second can be computed exactly and is finite for $Q\to 0$. Hence the 
singular part
of $J$ is given by
\begin{eqnarray}
J &\approx& \int_{-\infty}^{+\infty} dx 
   \Bigl[ Q^{-3/2} g_{-4}(x) + Q^{-1} g_{-3}(x) +\nonumber\\
&+& 
Q^{-1/2} g_{-2} (x) + 
          g_{-1}(x) - {g_{-1,\infty} x\over 1 + x^2}\Bigr] .
\end{eqnarray}
Using again the fact that the expansion (\ref{exphoverF})
is in powers of $x\sqrt{Q}$, we observe the $g_{2n}(x)$ is even under 
$x\to -x$, while $g_{2n+1}(x)$ is odd. We obtain finally
\begin{equation}
J = J_{-3/2} Q^{-3/2} + J_{-1/2} Q^{-1/2} + O(1),
\end{equation}
with 
\begin{eqnarray}
J_{-3/2} &=& \int_{-\infty}^{+\infty} dx\, g_{-4}(x),  \\
J_{-1/2} &=& \int_{-\infty}^{+\infty} dx\, g_{-2}(x).
\end{eqnarray}
If we now write 
$F(w) = w^2 + \sum_{n\ge 3} F_n w^n$, $h(w) = \sum_n h_n w^n$, we obtain
\begin{eqnarray}
J_{-3/2} &=& {\pi h_0\over2}, \\
J_{-1/2} &=& {\pi\over 16} (15 F_3^2 h_0 - 12 F_4 h_0 - 12 F_3 h_1 + 8 h_2).
\nonumber
\end{eqnarray}
From the expansion of $J$ we can easily derive the expansion of $I$:
\begin{equation}
I = I_{-1/2} Q^{-1/2} + I_0 + I_{1/2} Q^{1/2} + O(Q),
\end{equation}
with 
$I_{-1/2} = 2 J_{-3/2}$ and 
$I_{1/2} = - 2 J_{-1/2}$.

\section{The canonical form of the trap $\beta$-function}
 \label{app:betatrap}

We wish now to prove that the trap $\beta$-function can be rewritten
as in Eq.~(\ref{beta-trap}). As in App.~\ref{app:beta} we work
by perturbative induction. We assume that we have proved the result
at order $n-1$, i.e. that the $\beta$-function has the form
\begin{equation}
\beta_w(u,v,w) = 
   w T(u,v) +w^2 \sum_{k=n} H_{k-2}(u,v,w),
\end{equation}
where $T(0,0) = 1$ and $H_{k}(u,v,w)$ are homogeneous polynomials of order 
$k$, i.e. satisfy
\begin{equation}
H_{k}(\lambda u,\lambda v,\lambda w) = 
\lambda^k H_{k}(u,v,w).
\end{equation}
Then, we perform the change of variables 
\begin{equation}
z = w + w G_{n-1}(u,v,w),
\end{equation}
where $G_{k}(u,v,w)$ is homogeneous polynomial of degree $k$. 
If we now compute the $\beta$-function associated with $z$,
\begin{equation}
  \beta_z(u,v,z) = {dz\over dl},
\end{equation}
we obtain 
\begin{equation}
\beta_z = z T(u,v) + 
     z^2 \left[H_{n-2}(u,v,z) + {\partial G_{n-1}\over \partial z}\right]
\end{equation}
where we neglect terms of order $n+1$ in the variables.
Hence, if we define 
\begin{equation}
G_{n-1}(u,v,w) = -\int_0^w dx H_{n-2}(u,v,x),
\end{equation}
where the integral is performed at fixed $u$ and $v$,
we cancel all unwanted terms, proving the result.


\begin{thebibliography}{99}

\bibitem{KT-73} 
J. M. Kosterlitz and  D. J. Thouless,
J.\ Phys. C: Solid State {\bf 6},  1181 (1973)

\bibitem{B-72}
V. L. Berezinskii, Sov. Phys. JETP {\bf 34}, 610 (1972).

\bibitem{Kosterlitz-74}
J. M. Kosterlitz, J. Phys. C {\bf 7}, 1046 (1974).

\bibitem{JKKN-77}
J. V. Jos\'e, L. P. Kadanoff, S. Kirkpatrick, and D. R. Nelson,
Phys. Rev. B {\bf 16}, 1217 (1977).

\bibitem{MW-66}
N. D. Mermin and H. Wagner, Phys. Rev. Lett. {\bf 17}, 1133 (1966).

\bibitem{H-67}
P. C. Hohenberg, Phys. Rev. {\bf 158}, 383 (1967).

\bibitem{HKCBD-06} Z. Hadzibabic, P. Kr\"uger, M. Cheneau, B. Battelier,
and J. Dalibard, Nature {\bf 441}, 1118 (2006).

\bibitem{KHD-07}
P. Kr\"uger, Z. Hadzibabic, and J. Dalibard,
Phys. Rev. Lett. {\bf 99}, 040402 (2007).

\bibitem{HKCRD-08}
Z. Hadzibabic, P. Kr\"uger, M. Cheneau, S. P. Rath,
and J. Dalibard, New J. Phys. {\bf 10}, 045006 (2008).

\bibitem{CRRHP-09} P. Clad\'e, C. Ryu, A. Ramanathan, K. Helmerson, 
and W. D.  Phillips, Phys. Rev. Lett. {\bf 102}, 170401 (2009).

\bibitem{HZGC-10}
C.-L. Hung, X. Zhang, N. Gemelke, and C. Chin,
Nature {\bf 470}, 236 (2011).

\bibitem{GKMD-08} 
F. M. Gasparini, M. O. Kimball, K. P. Mooney, and M. Diaz-Avilla,
Rev. Mod. Phys. {\bf 80}, 1009 (2008).

\bibitem{RGBSN-81}
D. J. Resnick, J. C. Garland, J. T. Boyd, S. Shoemaker, and R. S. Newrock, 
Phys. Rev. Lett. {\bf 47}, 1542 (1981).

\bibitem{HP-97}
M. Hasenbusch and  K. Pinn,
J. Phys. A {\bf 30}, 63 (1997).

\bibitem{Hasenbusch-05}
M. Hasenbusch,  J. Phys. A {\bf 38}, 5869 (2005). 

\bibitem{KO-12}
Y. Komura and Y. Okabe,
J. Phys. Soc. Jpn. {\bf 81}, 113001 (2012).

\bibitem{BC-93-94}
P. Butera and M. Comi,
Phys. Rev. B {\bf 47}, 11969 (1993);
{\em ibid.} {\bf 50}, 3052 (1994).

\bibitem{KI-95}
R. Kenna and A. C. Irving, Phys. Lett. B {\bf 351}, 273 (1995).

\bibitem{KI-96}
R. Kenna and A. C. Irving,
Phys. Rev. B {\bf 53}, 11568 (1996).

\bibitem{CPRV-96} 
M. Campostrini, A. Pelissetto, P. Rossi, and E. Vicari,
Phys. Rev. B {\bf 54}, 7301 (1996).

\bibitem{Janke-97}
W. Janke, Phys. Rev. B {\bf 55}, 3580 (1997).

\bibitem{KI-97}
R. Kenna and A. C. Irving, Nucl. Phys. B {\bf 485}, 583 (1997).

\bibitem{JH-98}
A. Jaster and H. Hahn, Physica A {\bf 252}, 199 (1998). 

\bibitem{Balog-01}
J. Balog, J. Phys. A {\bf 34}, 5237 (2001).

\bibitem{BNNPSW-01}
J. Balog, M. Niedermaier, F. Niedermayer, A. Patrascioiu, E. Seiler, and P.
Weisz, Nucl. Phys. B {\bf 618}, 315 (2001).

\bibitem{CS-03}
S. Chandrasekharan and C.G. Strouthas, Phys. Rev. D {\bf 68}, 091502 (2003).

\bibitem{BP-08}
P. Butera and M. Pernici,
Physica A {\bf 387}, 6293 (2008).

\bibitem{Arisue-09}
H. Arisue, Phys. Rev. E {\bf 79}, 011107 (2009).

\bibitem{Baxter-82}
R.J. Baxter, {\em Exactly Solved Models in Statistical Mechanics},
(New York: Academic, 1982).

\bibitem{CPRV-96b}
M. Campostrini, A. Pelissetto, P. Rossi, and E. Vicari,
Nucl. Phys. B {\bf 459}, 207 (1996).

\bibitem{PV-98}
A. Pelissetto and E. Vicari,
Nucl. Phys. B {\bf 519}, 626 (1998);
Nucl. Phys. B {\bf 522}, 605 (1998).

\bibitem{BNNPSW-00}
J. Balog, M. Niedermaier, F. Niedermayer, A. Patrascioiu, E. Seiler, and P.
Weisz, Nucl. Phys. B {\bf 583}, 614 (2000).

\bibitem{PV-r}
A. Pelissetto and E. Vicari,
Phys. Rep. {\bf 368}, 549 (2002).


\bibitem{AGG-80}
D. J. Amit, Y. Y. Goldschmidt, and G. Grinstein,
J. Phys. A {\bf 13}, 585 (1980)

\bibitem{FBJ-73}
M. E. Fisher, M. N. Barber, and D. Jasnow,
Phys. Rev. A {\bf 8}, 1111 (1973).

\bibitem{Cardy-88}
J. Cardy, {\em Finite-Size Scaling}
(North Holland, Amsterdam, 1988).

\bibitem{CV-09} M. Campostrini and E. Vicari, Phys. Rev. Lett. {\bf 102},
240601 (2009); (E) {\bf 103}, 269901 (2009).

\bibitem{Wegner-76}
F. J. Wegner, The critical state, general aspects, in {\em 
Phase Transitions and Critical Phenomena}, Vol. 6,
edited by C. Domb and M. S. Green (Academic Press, London, 1976), p. 7. 

\bibitem{PHA-91}
V. Privman, P.C. Hohenberg, and A.A. Aharony, 
in {\em Phase Transitions and Critical Phenomena}, Vol.~14,
edited by C. Domb and J. L. Lebowitz
(Academic Press, London-San Diego, 1991).

\bibitem{Privman-90}
V. Privman (ed.), 
{\em Finite Size Scaling and Numerical Simulations of Statistical Systems}
(World Scientific, Singapore, 1990).

\bibitem{SS-99}
J. Salas and A. D. Sokal, e-print cond-mat/9904038v1;
J. Stat. Phys. {\bf 98}, 551 (2000).

\bibitem{ZJ-book} 
J. Zinn-Justin, \emph{Quantum Field Theory and Critical Phenomena},
fourth edition 
(Clarendon Press, Oxford, 2002).

\bibitem{BH-00}
J. Balog and A. Heged\~us,
J. Phys. A {\bf 33}, 6543 (2000)

\bibitem{footnotesbf} Note that the $\beta$-functions defined in this
paper differ by a sign from those reported in Ref.~\cite{BH-00}.The
reason is that here we define them by taking derivatives with respect
to a length scale $b$, while in Ref.~\cite{BH-00} the ``mass" $m=1/b$
is used.

\bibitem{APV-10} V. Alba, A. Pelissetto, and E. Vicari,
J. Stat. Mech. P03006 (2010).

\bibitem{Diehl-86} H.W. Diehl, {\em Field Theoretical Approach at
Surfaces in Phase Transitions and Critical Phenomena}, edited by
C. Domb and J.L. Lebowitz, Vol. 10 (Academic Press, London 1986)
p. 76.

\bibitem{Hasenbusch-12} M. Hasenbusch, Phys. Rev. B {\bf 85}, 174421
(2012).


\bibitem{Hasenbusch-08}
M. Hasenbusch,  J. Stat. Mech. P08003 (2008).

\bibitem{Hasenbusch-09}
M. Hasenbusch,  J. Stat. Mech. P02005 (2009).

\bibitem{HPV-05} 
M. Hasenbusch, A. Pelissetto, and E. Vicari, J. Stat. Mech.  P12002 (2005).

\bibitem{CWK-02} E. A. Cornell and C. E. Wieman, 
Rev.\ Mod.\ Phys.\ {\bf 74}, 875 (2002); 
N. Ketterle, Rev.\ Mod.\ Phys.\ {\bf 74}, 1131 (2002).

\bibitem{BDZ-08} 
I. Bloch, J. Dalibard, and W. Zwerger, 
Rev.\ Mod.\ Phys.\ {\bf 80}, 885 (2008).

\bibitem{CV-10}
M. Campostrini and E. Vicari,
Phys. Rev. A {\bf 81}, 023606 (2010).

\bibitem{JBCGZ-98}
D. Jaksch, C. Bruder, J. I. Cirac, C. W. Gardiner, and P. Zoller,
Phys. Rev. Lett. {\bf 81}, 3108 (1998).

\bibitem{FWGF-89}
M. P. A. Fisher, P. B. Weichman, G. Grinstein, and D. S. Fisher,
Phys. Rev. B {\bf 40}, 546 (1989).

\bibitem{CR-12}
J. Carrasquilla and M. Rigol, Phys. Rev. A {\bf 86}, 043629 (2012).

\bibitem{CV-11}
F. Crecchi and E. Vicari, Phys. Rev. A {\bf 83},  035602 (2011).

\bibitem{QSS-10} 
S. L. A.~de~Queiroz, R. R.~dos~Santos, and R. B.~Stinchcombe,
Phys. Rev. E {\bf 81}, 051122 (2010).

\bibitem{CTV-13}
G. Ceccarelli, C. Torrero, and E. Vicari, e-print arXiv:1211.6224.

\bibitem{CV-10b} M. Campostrini and E. Vicari, 
Phys. Rev. A {\bf 81}, 063614 (2010).

\bibitem{CTV-12}
G. Ceccarelli, C. Torrero, and E. Vicari, Phys. Rev. A {\bf 85}, 023616 (2012).

\bibitem{CT-12}
G. Ceccarelli and C. Torrero, Phys. Rev. A {\bf 85}, 053637 (2012).

\bibitem{CG-99} G. Cicogna and G. Gaeta, {\em Symmetry and
perturbation theory in nonlinear dynamics}, Lecture notes in Physics
(Springer, Berlin-Heidelberg, 1999), chapter 8.

\bibitem{Arnold}
V. I. Arnol'd, 
{\em Ordinary Differential Equations}
(Springer, Berlin-Heidelberg, 1992).




\end{thebibliography}
\end{document}